\documentclass[journal]{IEEEtran}
\usepackage{lineno}
\usepackage{cite}
\usepackage{hyperref}
\usepackage{amsmath,amssymb,amsfonts}
\usepackage{amsthm}
\DeclareMathOperator*{\argmax}{argmax}

\usepackage{graphicx}
\usepackage{textcomp}
\usepackage{float}
\usepackage{subfigure}
\usepackage{amsmath}
\usepackage{mathrsfs}
\usepackage{mathtools}
\usepackage{algorithm}
\usepackage{algorithmicx}
\usepackage{algpseudocode}
\usepackage{bm}
\usepackage{multirow}
\usepackage{array}
\usepackage{url}
\usepackage{xcolor}
\usepackage{subfigure}
\usepackage{fancyhdr}
\usepackage{mdwmath}
\usepackage{mdwtab}
\usepackage{caption}
\usepackage[justification=centering]{caption}
\usepackage{setspace}
\usepackage{algpseudocode}
\usepackage{dsfont}
\usepackage{bbm}
\usepackage{cases}
\usepackage{stfloats}
\newtheorem{remark}{Remark}
\theoremstyle{definition}
\newtheorem{theorem}{Theorem}

\newtheorem{lemma}{Lemma}

\newtheorem{corollary}{Corollary}

\makeatletter
\newcommand{\biggg}{\bBigg@{3}}
\newcommand{\Biggg}{\bBigg@{3.5}}
\makeatother
\begin{document}
	
	\title{Performance Analysis of Physical Layer Security: \\From Far-Field to Near-Field}
	\author{Boqun~Zhao,~\IEEEmembership{Graduate Student Member,~IEEE,}~Chongjun~Ouyang,~\IEEEmembership{Member,~IEEE,}\\Xingqi~Zhang,~\IEEEmembership{Senior Member,~IEEE,}
		and Yuanwei~Liu,~\IEEEmembership{Fellow,~IEEE}\vspace{-10pt}
		\thanks{An earlier version of this paper is to be presented in part at the IEEE International Conference on Communications, Montreal, Canada, June 2025 \cite{icc_2025}. This work was supported by the Natural Sciences and Engineering Research Council of Canada (NSERC), the Future Energy Systems (FES) - Canada First Research Excellence Fund (CFREF), and the Alberta Innovates. The work of Boqun Zhao was supported in part by China Scholarship Council. The work of Chongjun Ouyang was supported in part by the Marie Skłodowska-Curie Actions (MSCA) Postdoctoral Fellowship and in part by the U.K. Engineering and Physical Sciences Research Council (EPSRC) under Grant EP/Z003091/1. \textit{(Corresponding authors: Xingqi Zhang, Yuanwei Liu.)}}
		\thanks{B. Zhao and X. Zhang are with the Department of Electrical and Computer Engineering, University of Alberta, Edmonton AB, T6G 2R3, Canada (email: \{boqun1, xingqi.zhang\}@ualberta.ca).}
		\thanks{C. Ouyang is with the School of Electronic Engineering and Computer Science, Queen Mary University of London, London E1 4NS, U.K., and also with the School of Electrical and Electronic Engineering, The University of Manchester, Manchester, U.K. (e-mail: c.ouyang@qmul.ac.uk).}
		\thanks{Y. Liu is with the Department of Electrical and Electronic Engineering, The University of Hong Kong, Hong Kong (e-mail: yuanwei@hku.hk).}
	}
	
	\maketitle
	
	\begin{abstract}
		The secrecy performance in both near-field and far-field communications is analyzed using two fundamental metrics: the \emph{secrecy capacity} under a power constraint and the \emph{minimum power requirement} to achieve a specified secrecy rate target. 1) For the secrecy capacity, a closed-form expression is derived under a discrete-time memoryless setup. This expression is further analyzed under several far-field and near-field channel models, and the capacity scaling law is revealed by assuming an infinitely large transmit array and an infinitely high power. A novel concept of \emph{``depth of insecurity''} is proposed to evaluate the secrecy performance achieved by near-field beamfocusing. It is demonstrated that increasing the number of transmit antennas reduces this depth and thus improves the secrecy performance. 2) Regarding the minimum required power, a closed-form expression is derived and analyzed within far-field and near-field scenarios. Asymptotic analyses are performed by setting the number of transmit antennas to infinity to unveil the power scaling law. Numerical results are provided to demonstrate that: \romannumeral1) compared to far-field communications, near-field communications expand the areas where secure transmission is feasible, specifically when the eavesdropper is located in the same direction as the intended receiver; \romannumeral2) as the number of transmit antennas increases, neither the secrecy capacity nor the minimum required power scales or vanishes unboundedly, adhering to the principle of energy conservation.
	\end{abstract}
	
	
	\section{Introduction}
	Wireless channels, due to their inherent broadcast nature, expose transmitted signals to potentially insecure environments, making them susceptible to interception by eavesdroppers. This vulnerability underscores the critical importance of secrecy performance in wireless communications. Unlike traditional cryptography-based security methods in the upper layers (e.g., the network layer), secure channel coding offers a robust alternative by enabling secure transmissions at the physical layer, a concept known as physical layer security. This approach effectively bypasses the need for additional spectral resources and minimizes signaling overhead \cite{PLS_survey}. The basic model for protecting information at the physical layer is Wyner's wiretap channel \cite{wyner}, which also introduced the associated notion of secrecy capacity, measuring the supremum of the achievable secure coding rate. 
	
	Recent research trends have highlighted the advantages of multiple-antenna technologies in improving secrecy capacity through beamforming design \cite{PLS_survey,wu2018survey}. Intuitively, using a larger-aperture antenna array can provide a higher beamforming gain and, consequently, a higher secure transmission rate. With advancements in antenna manufacturing and to meet the stringent security demands of future networks, there is a growing trend for base stations (BSs) to employ extremely large-aperture antenna arrays and operate at significantly higher frequencies \cite{liu2023near,liu2024near,wangzhe_1,wangzhe_2}. However, this shift represents not just a quantitative increase in antenna size and carrier frequency but also a qualitative paradigm shift from conventional far-field communications (FFC) to near-field communications (NFC). Specifically, based on the criterion for differentiating between near-field and far-field regions---the \emph{Rayleigh distance} $\frac{2D^2}{\lambda}$, where $D$ represents the antenna aperture and $\lambda$ is the wavelength \cite{rayleighdis}---the near-field region is expected to expand substantially. For instance, an array with an aperture of $D = 0.5$ m operating at a frequency of 60 GHz would encompass a near-field region extending up to $100$ meters.
	
	\begin{figure}[!t]
		\centering
		\setlength{\abovecaptionskip}{3pt}
		\includegraphics[height=0.13\textwidth]{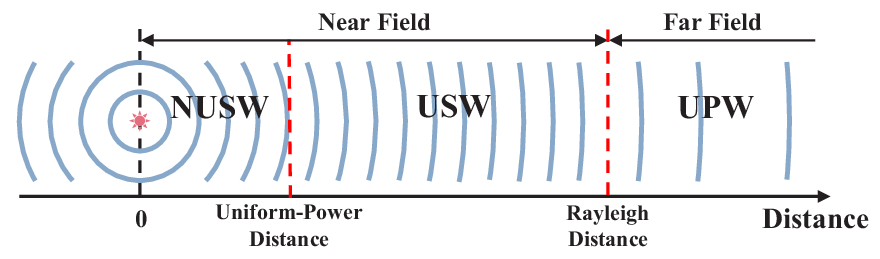}
		\caption{Illustration of the spherical waves with distance.}
		\label{wave_model}
		\vspace{-8pt}
	\end{figure}
	
	Electromagnetic (EM) waves exhibit distinct propagation characteristics in the near-field region compared to the far field. The space surrounding an antenna array can be divided into three regions, delineated by two specific distances: the \emph{Rayleigh distance} and the \emph{uniform-power distance} \cite{liu2023near,ouyang2024primer}. As mentioned earlier, the Rayleigh distance serves as the boundary between the near-field and far-field regions. When the signal’s propagation distance exceeds the Rayleigh distance, i.e., in the far-field region, EM waves are accurately modeled as planar waves, resulting in a linearly varying phase shifts of the signals. In contrast, if the propagation distance is shorter than the Rayleigh distance, i.e., in the near-field region, a spherical-wave-based channel model is necessary to accurately represent the non-linear varying of signal phase shifts. Additionally, within the near-field region, the uniform-power distance further distinguishes regions of uniform and non-uniform signal amplitude. Consequently, based on the Rayleigh and uniform-power distances, three channel models have been identified to accommodate various signal propagation distances with increasing accuracy: the uniform planar-wave (UPW) model, the uniform spherical-wave (USW) model, and the non-uniform spherical-wave (NUSW) model, as illustrated in {\figurename} {\ref{wave_model}}.

	\subsection{Prior Works}
		The secrecy performance for FFC, where the UPW model is applied, has been widely studied; see \cite{PLS_survey,wu2018survey,lv2024safeguarding,covert_1,covert_2}, and related references. However, research on near-field physical layer security is still in its early stages. The authors in \cite{nf_pls_1} utilized the USW model to study a near-field secure directional modulation system and propose a secure precoding algorithm. The work in \cite{nf_pls_2} leveraged the NUSW model to develop a two-stage algorithm that maximizes the near-field secrecy capacity via jointly optimizing unit-modulus phase shifters and baseband digital beamformers. The authors of \cite{nf_pls_3} introduced analog beamfocusing techniques using true-time delayers and phase shifters to effectively mitigate beamsplit and exploit near-field advantages, significantly improving physical layer security in near-field wideband communications. Meanwhile, the study in \cite{nf_pls_4} proposed an efficient low-complexity strategy for designing the beamforming with artificial noise to ensure near-field secure transmission with multiple legitimate users. 
		
		All the aforementioned works focused on secure beamforming design in NFC, while only a few studies have concentrated on the fundamental analysis of secrecy performance. For example, \cite{performance_1} explored physical layer security in extra-large multiple-input multiple-output systems. \cite{performance_2} investigated the enhancement of physical layer security through near-field beamforming with an extremely large uniform planar array (UPA), showing that near-field beamfocusing can improve secrecy capacity and jamming rejection. However, the performance analyses in both \cite{performance_1} and \cite{performance_2} rely primarily on numerical simulations, leaving a gap in providing deeper insights and intuitive comparisons between near-field and far-field secrecy performance, which could be better understood through analytical results.
	
	\subsection{Motivations \& Contributions}
	In light of the above discussion, the UPW model, commonly utilized in far-field scenarios, is inapplicable for NFC. Therefore, it becomes imperative to reevaluate the performance for physical layer security from a near-field perspective. By examining the propagation differences between near-field and far-field channel models and their impact on secure transmission, we aim to provide a straightforward comparison of secrecy performance between NFC and FFC, which motivates this work.
	
	This article analyzes the performance of physical layer security across commonly used far-field and near-field channel models, namely the UPW, USW, and NUSW models. We derive closed-form expressions for both the secrecy capacity and the minimum required power. Besides, asymptotic analysis is performed to provide deeper insights. Our main contributions are summarized as follows.
	
	\begin{itemize}
		\item We investigate secure transmission scenarios where a large‐aperture array communicates with a legitimate user while being eavesdropped by another user. We provide detailed analysis of various channel models applicable across different ranges in both near-field and far-field regions, including the UPW, USW and NUSW models. In this context, we introduce two key metrics to measure secrecy performance: the \emph{secrecy capacity} under a power constraint and the \emph{minimum power requirement} needed to achieve a target secrecy rate.
		\item We derive a closed-form expression for the secrecy capacity by considering a discrete-time memoryless Gaussian channel. This expression is then analyzed under various far-field and near-field channel models to facilitate comparison. To glean further insights, we investigate the \emph{capacity scaling law} by considering scenarios with an infinitely large transmit array and infinitely high power. Additionally, we introduce a novel metric, termed \emph{``depth of insecurity''}, to quantify the impact of near-field beamfocusing and array aperture on secrecy performance.
		\item We also derive a closed-form expression for the minimum required transmit power to achieve a specific target secrecy rate. An asymptotic analysis is conducted to unveil the power scaling law by considering an infinitely large number of transmit antennas within each channel model. This analysis highlights the superiority of the NUSW model over the other models in terms of power scaling and energy conservation.
		\item We present numerical results demonstrating that NFC can reduce the areas where secure transmission is compromised by the presence of an eavesdropper. Additionally, the \emph{``depth of insecurity''} decreases as the number of antennas increases. Furthermore, unlike the UPW/USW models, the secrecy capacity does not grow indefinitely, nor does the minimum required power diminish without bounds as the number of transmit antennas increases under the NUSW model. This highlights the superior accuracy and robustness of the NUSW model.         
	\end{itemize}
	\subsection{Organization and Notations}
	The rest of this paper is organized as follows. Section \ref{system} presents the secure transmission framework, detailing the near-field and far-field channel models, and introducing the key performance metrics. Section \ref{capacity} derives the secrecy capacity and introduces the concept of \emph{``depth of insecurity''}. Then, Section \ref{mini_Power} investigates the minimum required transmit power across various channel models. Section \ref{numerical} provides numerical results to validate the theoretical insights. Finally, Section \ref{conclusion} concludes the paper.
	\subsubsection*{Notations}
	Throughout this paper, scalars, vectors, and matrices are denoted by non-bold, bold lower-case, and bold upper-case letters, respectively. For the matrix $\mathbf{A}$, ${\mathbf{A}}^{\mathsf{T}}$ and ${\mathbf{A}}^{\mathsf{H}}$ denote the transpose and transpose conjugate of $\mathbf{A}$, respectively. For the square matrix $\mathbf{B}$, $\det(\mathbf{B})$ denotes the determinant of $\mathbf{B}$. The notations $\lvert a\rvert$ and $\lVert \mathbf{a} \rVert$ denote the magnitude and norm of scalar $a$ and vector $\mathbf{a}$, respectively. The identity matrix and zero matrix are represented by $\mathbf{I}$ and $\mathbf{0}$, respectively.  The set $\mathbbmss{C}$ stand for the complex space, and notation $\mathbbmss{E}\{\cdot\}$ represents mathematical expectation. The notation $f(x)=\mathcal{O}\left(g(x)\right)$ means that $\lim\sup_{x\rightarrow\infty}\frac{\lvert f(x)\rvert}{\lvert g(x)\rvert}<\infty$. The ceiling operator is shown by $\lceil\cdot\rceil$, and $[N]$ represents the integer set $\{1,\ldots,N\}$. And we generally reserve the symbols $I(\cdot)$ for mutual information and $H(\cdot)$ for entropy. Finally, ${\mathcal{CN}}({\bm\mu},\mathbf{X})$ is used to denote the circularly-symmetric complex Gaussian distribution with mean $\bm\mu$ and covariance matrix $\mathbf{X}$.
	\section{System Model}\label{system}
\begin{figure}[!t]
		\centering
		\setlength{\abovecaptionskip}{3pt}
		\includegraphics[height=0.29\textwidth]{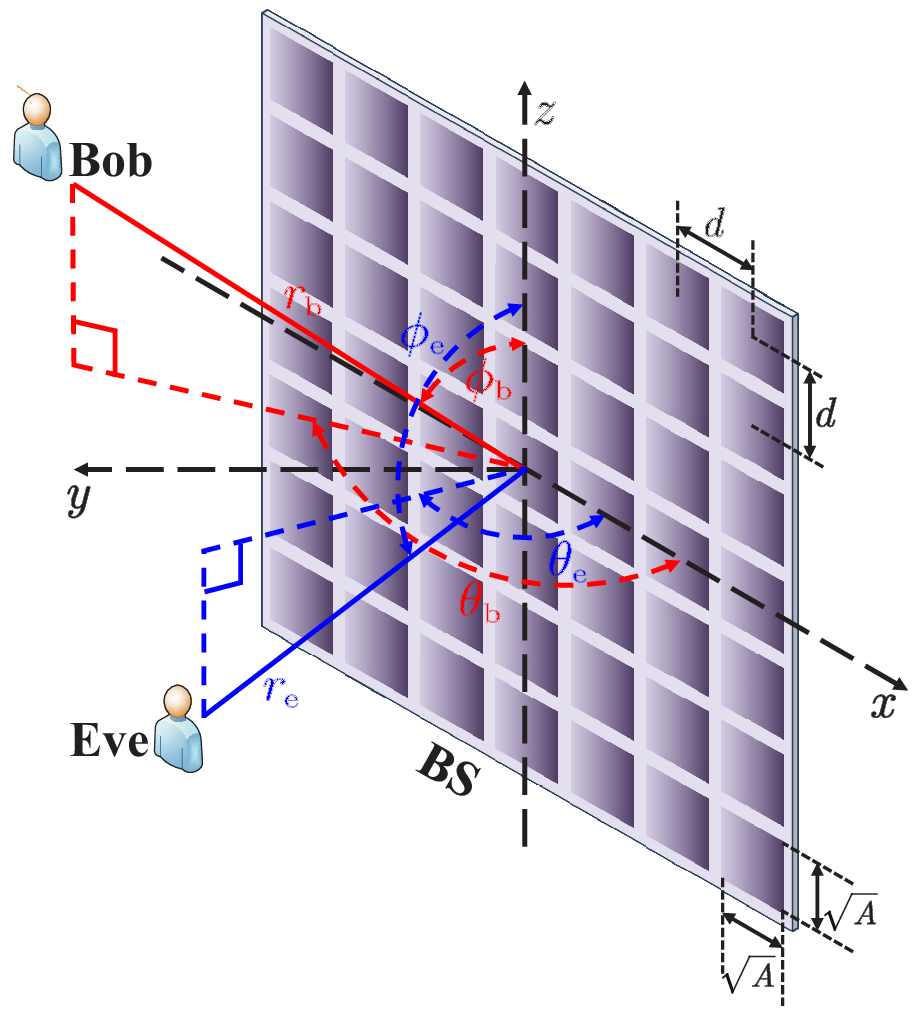}
		\caption{System layout of the wiretap channel.}
		\label{LoS_3D_Model}
		\vspace{-3pt}
	\end{figure}

	We consider a multiple-input single-output (MISO) wiretap channel, which comprises a base station (BS), a legitimate user (Bob), and an eavesdropper (Eve), as depicted in {\figurename} {\ref{LoS_3D_Model}}. The BS is equipped with a large-aperture UPA containing $M\gg1$ antenna elements, while both users are single-antenna devices \footnote{The two-user wiretap channel considered in this work is a basic model for secret communications, which has been extensively studied in prior research, e.g., \cite{PLS_1,khisti2010secure,khisti2010secure_2,nf_pls_1,nf_pls_2,nf_pls_3,performance_1,performance_2}. This model provides valuable insights into the fundamental limits of physical layer security and has facilitated significant theoretical advancements in secure communications, as demonstrated in \cite{leung1978gaussian,khisti2010secure,khisti2010secure_2,PLS_1,ly2010mimo,wyner}. Since this research represents an initial exploration into the fundamental performance limits of near-field physical layer security, we focus on this basic model to derive deeper insights into the system. Future studies will extend this analysis to more complex scenarios with multiple users.}. The UPA is placed on the $x$-$z$ plane and centered at the origin. Here, $M=M_{x}M_{z}$, where $M_{x}$ and $M_{z}$ denote the number of array elements along the $x$- and $z$-axes, respectively. For brevity, we assume that $M_{x}$ and $M_{z}$ are odd numbers with $M_x=2\tilde{M}_x+1$ and $M_z=2\tilde{M}_z+1$. The physical dimensions of each BS array element along the $x$- and $z$-axes are denoted by $\sqrt{A}$, and the inter-element distance is $d$, where $d\geq\sqrt{A}$. 
	
	A discrete-time memoryless and Gaussian channel setting is considered, where the BS intends to transmit a \emph{confidential message} $W$ at a rate ${\mathsf{R}}$ in bits per channel use to Bob over a transmission interval of length $N$, i.e., $N$ channel uses, such that it is kept secret from Eve. The signals observed at Bob (node $\mathsf{b}$) and Eve (node $\mathsf{e}$), respectively, are, for each time interval $n\in[N]$,
	\begin{subequations}\label{Received_Signal_ABE}
		\begin{align}
			y_{\mathsf{b}}(n)={\mathbf{h}}_{\mathsf{b}}^{\mathsf{H}}{\mathbf{x}}(n)+z_{\mathsf{b}}(n),\label{Received_Signal_Bob}\\
			y_{\mathsf{e}}(n)={\mathbf{h}}_{\mathsf{e}}^{\mathsf{H}}{\mathbf{x}}(n)+z_{\mathsf{e}}(n),\label{Received_Signal_Eve}
		\end{align}
	\end{subequations}
	where ${\mathbf{x}}(n)\in{\mathbbmss{C}}^{M\times1}$ is the transmitted signal vector, ${\mathbf{h}}_{\mathsf{b}}\in{\mathbbmss{C}}^{M\times1}$ and ${\mathbf{h}}_{\mathsf{e}}\in{\mathbbmss{C}}^{M\times1}$ denote the Bob-to-BS and Eve-to-BS channel vectors, respectively, and $z_{\mathsf{b}}(n)\sim{\mathcal{CN}}(0,\sigma_{\mathsf{b}}^2)$ and $z_{\mathsf{e}}(n)\sim{\mathcal{CN}}(0,\sigma_{\mathsf{e}}^2)$ are circularly symmetric complex-valued Gaussian noises with $\sigma_{\mathsf{b}}^2$ and $\sigma_{\mathsf{e}}^2$ being the noise powers. Moreover, the noises are independently sampled for each $n\in[N]$, and the input satisfies a power constraint of $P$, i.e., ${\mathbbmss{E}}\{\lVert{\mathbf{x}}\rVert^2\}\leq P$. 
	\subsection{Secure Transmission}\label{Section: System Model: Secure Transmission}
	The message $W$ is uniformly distributed over the index set ${\mathcal{W}}_N\triangleq\{1,\ldots,2^{\lceil N{\mathsf{R}}\rceil}\}$. To achieve secure transmission, the BS employs a channel encoding function $f_{\mathsf{en}}:{\mathcal{W}}_N\mapsto{\mathbbmss{C}}^{M\times N}$ to map the confidential message $W$ to the transmitted vector sequence $[{\mathbf{x}}(1),\ldots,{\mathbf{x}}(N)]\in{\mathbbmss{C}}^{M\times N}$, which is directed towards Bob while being overheard by Eve. Bob then uses a decoding function $f_{\mathsf{de}}:{\mathbbmss{C}}^{1\times N}\mapsto{\mathcal{W}}_N$ to make an estimate $\hat{W}$ of $W$ from the received output ${\mathbf{y}}_{\mathsf{b}}^{\mathsf{T}}=[y_{\mathsf{b}}(1),\ldots,y_{\mathsf{b}}(N)]\in{\mathbbmss{C}}^{1\times N}$ from his channel, incurring a block error rate $\varepsilon _N=\Pr(W\ne \hat{W})$. The confidential message $W$ is also the input to the eavesdropper's channel, and Eve has an average residual uncertainty $H(W|{\mathbf{y}}_{\mathsf{e}})$ after observing the output ${\mathbf{y}}_{\mathsf{e}}^{\mathsf{T}}=[y_{\mathsf{e}}(1),\ldots,y_{\mathsf{e}}(N)]$. We define the fractional equivocation of Eve as $\Delta_N\triangleq\frac{H(W|{\mathbf{y}}_{\mathsf{e}})}{H(W)}$, and the rate of secure transmission as $\frac{H(W)}{N}\triangleq\mathsf{R}$.
	\subsubsection*{Secrecy Capacity}
	A secrecy rate $\mathsf{R}$ is achievable if there exists a secrecy encoder-decoder tuple $({\mathcal{W}}_N,f_{\mathsf{en}},f_{\mathsf{de}})$ such that the equivocation $\Delta_N\rightarrow1$ and the probability of error $\varepsilon _N\rightarrow0$ as $N\rightarrow\infty$. The secrecy capacity is the supremum of all achievable secrecy rates.
	
	It is worth noting that $\Delta_N\rightarrow1$ is equivalent to $H(W|{\mathbf{y}}_{\mathsf{e}})\rightarrow H(W)$ or $I(W;{\mathbf{y}}_{\mathsf{e}})=H(W)-H(W|{\mathbf{y}}_{\mathsf{e}})\rightarrow0$, indicating that Eve cannot recover any information regarding the confidential message $W$.
	\subsubsection*{Capacity-Achieving Schemes}
	As proved in \cite{khisti2010secure}, a \emph{beamforming} strategy is \emph{capacity-achieving}. More specifically, the BS first encodes the confidential message $W$ into the codeword $[s(1),\ldots,s(N)]$ using a code for the scalar Gaussian wiretap channel \cite{leung1978gaussian}. The encoded symbols are distributed as Gaussian with zero mean and unit variance, i.e., $s(n)\sim{\mathcal{CN}}(0,1)$ for $n\in[N]$. The BS then maps the encoded symbols in the time interval $n\in[N]$, i.e., $s(n)$, into the transmit signal ${\mathbf{x}}(n)$ via a linear digital beamformer ${\mathbf{w}}\in{\mathbbmss{C}}^{M\times1}$. As a result, the transmit signal can be expressed as follows:
	\begin{equation}\label{Transmit_Signal}
		{\mathbf{x}}(n)={\mathbf{w}}s(n),
	\end{equation}
	where $\lVert{\mathbf{w}}\rVert^2\leq P$ is the power budget. Given $\mathbf{w}$, the effective channel and receive signal-to-noise ratio (SNR) of node $k$ are given by ${\mathbf{h}}_k^{\mathsf{H}}{\mathbf{w}}$ and $\gamma_k=\frac{1}{\sigma_k^2}\lvert{\mathbf{h}}_k^{\mathsf{H}}{\mathbf{w}}\rvert^2$, respectively. The resulting secrecy capacity equals that of a scalar Gaussian wiretap channel, which can be written as follows \cite{leung1978gaussian}: 
	\begin{equation}\label{Secrecy_Channel_Capacity}
		{\mathsf{C}}_{\mathsf{w}}
		=\max\left\{\log_2\left(\frac{1+\gamma_{\mathsf{b}}}{1+\gamma_{\mathsf{e}}}\right),0\right\}.
	\end{equation}
	The secrecy capacity of our considered system, as defined in \eqref{Received_Signal_ABE}, is thus given by the supremum of $\{{\mathsf{C}}_{\mathsf{w}}|\lVert{\mathbf{w}}\rVert^2\leq P\}$.
	\subsubsection{Secrecy Channel Capacity}
	The above arguments imply that the secrecy capacity can be expressed as follows:
	\begin{align}\label{Secrecy_Channel_Capacity_2}
		{\mathsf{C}}&=\max_{\lVert{\mathbf{w}}\rVert^2\leq P}{\mathsf{C}}_{\mathsf{w}}\nonumber\\
		&=\max_{\lVert{\mathbf{w}}\rVert^2\leq P}\left(\max\left\{\log_2\left(\frac{1+\gamma_{\mathsf{b}}}{1+\gamma_{\mathsf{e}}}\right),0\right\}\right).
	\end{align}
	The secrecy capacity represents the supremum of the achievable secrecy rates subject to a power budget. 
	\subsubsection{Minimum Required Transmit Power}
	Another metric of interest in this work is the minimum required transmit power to guarantee a target secrecy rate ${\mathsf{R}}_0>0$. This is given by
	\begin{equation}\label{Secrecy_Transmit_Power}
		{\mathsf{P}}=\min_{{\mathsf{C}}_{\mathbf{w}}\geq {\mathsf{R}}_{0}}\lVert{\mathbf{w}}\rVert^2
		=\min\nolimits_{\log_2\left({\frac{1+\gamma_{\mathsf{b}}}{1+\gamma_{\mathsf{e}}}}\right)\geq {\mathsf{R}}_{0}}\lVert{\mathbf{w}}\rVert^2.
	\end{equation}
	
	Both the secrecy capacity and the minimum required power are essential metrics for evaluating the performance of secure transmission systems. The secrecy capacity represents the maximum transmission rate at which data can be securely transmitted, while the minimum required transmit power indicates the least amount of power needed to achieve a specified level of security. This manuscript aims to analyze these two metrics by examining several representative far-field and near-field channel models, which will be detailed later.
	\subsection{Channel Model}\label{Section: System Model: Near-Field Channel Model}
	All channel gains are fixed throughout the entire transmission period, and each node $k\in\{{\mathsf{b}},{\mathsf{e}}\}$ is assumed to possess complete channel state information (CSI) about its effective channel. It is further assumed that Eve is a registered user, and hence, during the channel training phase, the BS acquires the CSI of both Eve and Bob. Under these circumstances, Eve is expected to receive the \emph{common messages} broadcasted in the network but should remain \emph{uninformed} about the \emph{confidential messages} intended for Bob \footnote{This assumption is widely used in the physical layer security literature because many key secrecy performance metrics, such as secrecy channel capacity \cite{leung1978gaussian}, and foundational results, including secrecy channel coding theorems \cite{leung1978gaussian,khisti2010secure,khisti2010secure_2,wyner}, are derived based on it. Consequently, numerous studies on near-field physical layer security have also adopted this classical model; see \cite{nf_pls_2} and related references. For scenarios where Eve operates as a malicious node with unknown CSI and multiple antennas, techniques like artificial noise can be employed to enhance security; detailed discussions can be found in \cite{wang2012distributed,wang2020intelligent,cheng2024movable}. However, addressing such cases falls outside the scope of this work and will be explored in future research.}.
	
	Since near-field channels are sparsely-scattered and dominated by line-of-sight (LoS) propagation \cite{liu2023near}, we consider pure LoS propagation scenarios for a theoretical exploration of fundamental secrecy performance limits and variations as the propagation transitions from far-field to near-field. Following our description concerning {\figurename} {\ref{LoS_3D_Model}}, the central location of the $(m_x,m_z)$th BS element is denoted by ${\mathbf{s}}_{m_x,m_z}=[m_xd,0,m_zd]^{\mathsf{T}}$, where $m_x\in{\mathcal{M}}_x\triangleq\{0,\pm1,\ldots,\pm\tilde{M}_x\}$ and $m_z\in{\mathcal{M}}_z\triangleq\{0,\pm1,\ldots,\pm\tilde{M}_z\}$. Let $r_k$ denote the distance from the center of the antenna array to each node $k\in\{{\mathsf{b}},{\mathsf{e}}\}$, and $\theta_k\in(0,\pi)$ and $\phi_k\in(0,\pi)$ denote the azimuth and elevation angles, respectively. Thus, the location of node $k$ can be expressed as ${\mathbf{r}}_k=[r_k\Phi_k,r_k\Psi_k,r_k\Omega_k]^{\mathsf{T}}$, where $\Phi_k\triangleq\sin{\phi_k}\cos{\theta_k}$, $\Psi_k\triangleq\sin\phi_k\sin\theta_k$, and $\Omega_k\triangleq\cos{\phi_k}$. Moreover, we assume that both Bob and Eve are equipped with a single hypothetical isotropic antenna to receive incoming signals.
	\subsubsection{NUSW Model}
	When the users are located very close to the BS, i.e., the distance $r_k$ is smaller than the \emph{uniform-power distance} shown in {\figurename} {\ref{wave_model}}, it becomes imperative to differentiate both the power and phase of different elements for modeling the channel between the user and the BS array. Specifically, the distance between node $k$ and the center of the $(m_x,m_z)$th BS array element is given by
		\begin{align}
			&r_{m_x,m_z,k}=\lVert{\mathbf{r}}_k-{\mathbf{s}}_{m_x,m_z}\rVert\nonumber\\
			&=r_k\sqrt{1-2m_x\varepsilon_k\Phi_k-2m_z\varepsilon_k\Omega_k+(m_x^2+m_z^2)\varepsilon_k^2},\label{distance}
		\end{align}
	where $\varepsilon_k=\frac{d}{r_k}$. Note that $r_k=r_{0,0,k}$ and, since the array element separation $d$ is typically on the order of wavelength, in practice, we have $\varepsilon_k\ll1$. The channel response from node $k$ to the $(m_x,m_z)$th antenna element of the UPA can be written as $h_{m_x,m_z,k}=\sqrt{g_{m_x,m_z,k}}{\rm{e}}^{-{\rm{j}}\phi_{m_x,m_z,k}}$, where $g_{m_x,m_z,k}$ and $\phi_{m_x,m_z,k}$ are the channel power and phase, respectively. Building on our previous work \cite{ouyang2024impact}, we model the channel power $g_{m_x,m_z,k}$ as follows:
		\begin{align}
			g_{m_x,m_z,k}=\int_{{\mathcal{S}}_{m_x,m_z}}
			{\frac{\lvert{\mathbf{e}}_y^{\mathsf{T}}({\mathbf{s}}-{\mathbf{r}}_k)\rvert}{\lVert{\mathbf{r}}_k-{\mathbf{s}}\rVert}}{\mathsf{g}}(\mathbf{r}_k,{\mathbf{s}}){\rm{d}}{\mathbf{s}}.
		\end{align}
		where ${\mathcal{S}}_{m_x,m_z}\triangleq[m_xd-{\sqrt{A}}/{2},m_xd+{\sqrt{A}}/{2}]\times[m_zd-{\sqrt{A}}/{2},m_zd+{\sqrt{A}}/{2}]$ denotes the surface region of the $(m_x,m_z)$th array element. The term $\frac{{\mathbf{e}}_y^{\mathsf{T}}({\mathbf{r}}_k-{\mathbf{s}})}{\lVert{\mathbf{r}}_k-{\mathbf{s}}\rVert}$ models the influence of the projected aperture of each array element, which is reflected by the projection of the UPA normal vector ${\mathbf{e}}_y=[0,1,0]^{\mathsf{T}}$ onto the wave propagation direction at each local point $\mathbf{s}$ \cite{liu2023near}. It is applied to individual array elements that explicitly accounts for their size. Furthermore, ${\mathsf{g}}(\mathbf{r},{\mathbf{s}})$ models the effect of free-space EM propagation, which is given by \cite{ouyang2024impact}
		\begin{equation}\label{Green_Function_Full_Version}
			{\mathsf{g}}(\mathbf{r},{\mathbf{s}})=\frac{1}{{4\pi} \lVert{\mathbf{r}}-{\mathbf{s}}\rVert^2}
			\bigg(1-\frac{1}{k_0^2\lVert{\mathbf{r}}-{\mathbf{s}}\rVert^2}+\frac{1}{k_0^4\lVert{\mathbf{r}}-{\mathbf{s}}\rVert^4}\bigg),
		\end{equation}
		with $k_0=\frac{2\pi}{\lambda}$ being the wavenumber.
	
	Given the small antenna size compared to the propagation distance between the user and the antenna elements, i.e., $\sqrt{A}\ll\lVert{\mathbf{r}}_k-{\mathbf{s}}_{m_x,m_z}\rVert$, the variation in the channel response across an individual antenna element can be considered negligible. It follows from the fact of $\int_{{\mathcal{S}}_{m_x,m_z}}{\rm{d}}{\mathbf{s}}=A$ that
	\begin{align}
		g_{m_x,m_z,k}&\approx 
		\frac{\lvert{\mathbf{e}}_y^{\mathsf{T}}({\mathbf{s}}_{m_x,m_z}-{\mathbf{r}}_k)\rvert}{\lVert{\mathbf{r}}_k-{\mathbf{s}}_{m_x,m_z}\rVert}{\mathsf{g}}(\mathbf{r}_k,{\mathbf{s}}_{m_x,m_z})\int_{{\mathcal{S}}_{m_x,m_z}}{\rm{d}}{\mathbf{s}}\notag\\
		&=A\frac{\lvert{\mathbf{e}}_y^{\mathsf{T}}({\mathbf{s}}_{m_x,m_z}-{\mathbf{r}}_k)\rvert}{r_{m_x,m_z,k}}{\mathsf{g}}(\mathbf{r}_k,{\mathbf{s}}_{m_x,m_z}).\label{SPD_Electric_Field_Antenna_Element}
	\end{align}
	Furthermore, applying this approximation to the phase component, we obtain $\phi_{m_x,m_z,k}\approx\frac{2\pi}{\lambda}r_{m_x,m_z,k}$. The function ${\mathsf{g}}(\mathbf{r},{\mathbf{s}})$ comprises three terms: the first term corresponds to the radiating near-field and far-field regions, while the remaining two terms correspond to the reactive near-field region. Note that $1-\frac{1}{k_0^2\lVert{\mathbf{r}}-{\mathbf{s}}\rVert^2}+\frac{1}{k_0^4\lVert{\mathbf{r}}-{\mathbf{s}}\rVert^4}\approx0.97$ at distance $\lVert{\mathbf{r}}-{\mathbf{s}}\rVert=\lambda$. Hence, when considering practical NFC systems with $r_k\gg \lambda$, the last two terms in \eqref{Green_Function_Full_Version} can be neglected. Taken together, the channel response vectors can be modeled as follows:
	\begin{align}\label{channel_NF}
		\mathbf{h}_k= \left[\sqrt{\frac{Ar_k\Psi_k}{4\pi r_{m_x,m_z,k}^3}}\mathrm{e}^{-\mathrm{j}\frac{2\pi}{\lambda}r_{m_x,m_z,k}}\right]_{\forall m_x,m_z}.
	\end{align} 
	This model is known as the NUSW model, which is accurate but complex. 
	\subsubsection{USW Model}
	When the near-field user is located beyond the uniform-power distance, it is appropriate to simplify the model by assuming that the channel power disparity across each link becomes negligible, resulting in the USW model:
	\begin{align}
		\mathbf{h}_k=\sqrt{\frac{A\Psi _k}{4\pi r_{k}^{2}}}\left[ \mathrm{e}^{-\mathrm{j}\frac{2\pi}{\lambda}r_{m_x,m_z,k}} \right] _{\forall m_x,m_z}. 
	\end{align}
	Since the uniform-power distance is typically much smaller than the Rayleigh distance \cite{liu2023near}, the USW is effective for the majority of the area within the near-field region. Furthermore, as discussed in \cite{liu2023near}, when the communication distance exceeds the Fresnel distance $0.5\sqrt{\frac{D^3}{\lambda}}$ \cite{fresnel}, \eqref{distance} can be approximated as follows:
	\begin{equation}\label{distance_appro}
		\begin{split}
			&r_{m_x,m_z,k}\approx r_k\Big(1-\varepsilon _k(m_x\Phi _k+m_z\Omega _k)\\
			&+\frac{\varepsilon _{k}^{2}}{2}\big(m_{x}^{2}(1-\Phi _{k}^{2})+m_{z}^{2}(1-\Omega _{k}^{2})\big)\Big)\triangleq \tilde{r}_{m_x,m_z,k},    
		\end{split}   
	\end{equation}
	which is obtained by applying the Maclaurin series expansion $\sqrt{1+x}\approx 1+\frac{x}{2}-\frac{x^2}{8}$ and omitting the bilinear term. Notably, the Fresnel distance is much smaller than the Rayleigh distance \cite{fresnel,liu2023near}, which indicates that the above approximation is sufficiently accurate for the USW model. Therefore, the USW channel model can be approximated as follows:
	\begin{equation}\label{usw_model}
		\begin{split}
			\mathbf{h}_k\approx\sqrt{\frac{A\Psi _k}{4\pi r_{k}^{2}}}\left[ \mathrm{e}^{-\mathrm{j}\frac{2\pi}{\lambda}\tilde{r}_{m_x,m_z,k}} \right] _{\forall m_x,m_z}.
		\end{split}
	\end{equation}
	As both the uniform-power distance and Fresnel distance are close to the BS, the approximated USW model can adequately address most scenarios in NFC.
	\subsubsection{UPW Model}
	By positioning the users beyond the Rayleigh distance, the quadratic terms in \eqref{distance_appro} can be omitted, simplifying the USW-based near-field channel model into the UPW-based far-field model:
	\begin{align}\label{channel_FF}
		\mathbf{h}_k=\sqrt{\frac{A\Psi _k}{4\pi r_{k}^{2}}}\left[\mathrm{e} ^{-\mathrm{j}\frac{2\pi r_k}{\lambda}(1-m_x\varepsilon  _k\Phi _k-m_z\varepsilon  _k\Omega _k)} \right] _{\forall m_x,m_z}.    
	\end{align}
	The UPW model assumes that the angles of the links between each array element and node $k$ are approximately identical, leading to linearly varying phase shifts.
	
	In the following sections, we aim to analyze \eqref{Secrecy_Channel_Capacity_2} and \eqref{Secrecy_Transmit_Power} under the three models discussed above.
	\section{Analysis of the Secrecy Capacity}\label{capacity}
	In this section, we analyze the secrecy capacity and compare the performance between far-field and near-field scenarios.
	\subsection{Secrecy Capacity Characterization}
	According to the definition in \eqref{Secrecy_Channel_Capacity_2}, the optimal beamformer $\mathbf{w}^{\star}$ that achieves the secrecy capacity satisfies
	\begin{equation}\label{Optimal_Precoding}
		{\mathbf{w}^{\star}}=\argmax\nolimits_{\lVert{\mathbf{w}}\rVert^2\leq P}\frac{1+{\sigma_{\mathsf{b}}^{-2}}\lvert{\mathbf{h}}_{\mathsf{b}}^{\mathsf{H}}{\mathbf{w}}\rvert^2}{1+{\sigma_{\mathsf{e}}^{-2}}\lvert{\mathbf{h}}_{\mathsf{e}}^{\mathsf{H}}{\mathbf{w}}\rvert^2}.
	\end{equation}
	Based on the monotonicity of the function $f(x)=\frac{1+ax}{1+bx}$ for $x>0$, $a>0$, and $b>0$, it holds that $\frac{1+{\sigma_{\mathsf{b}}^{-2}}\lvert{\mathbf{h}}_{\mathsf{b}}^{\mathsf{H}}{\mathbf{w}}\rvert^2}{1+{\sigma_{\mathsf{e}}^{-2}}\lvert{\mathbf{h}}_{\mathsf{e}}^{\mathsf{H}}{\mathbf{w}}\rvert^2}$ is maximized when $\lVert{\mathbf w}\rVert^2=P$. Define $\overline{\gamma}_{k}\triangleq\frac{P}{\sigma_{k}^2}$ for ${k}\in\{{\mathsf b},{\mathsf e}\}$. Then, we rewrite problem \eqref{Optimal_Precoding} equivalently as follows:
	\begin{equation}\label{Optimal_Precoding_Trans}
		{\mathbf{v}^{\star}}=\argmax\nolimits_{\lVert{\mathbf{v}}\rVert^2=1}\left(\frac{1+{\overline{\gamma}_{\mathsf{b}}}\lvert{\mathbf{h}}_{\mathsf{b}}^{\mathsf{H}}{\mathbf{v}}\rvert^2}{1+
			{\overline{\gamma}_{\mathsf{e}}}\lvert{\mathbf{h}}_{\mathsf{e}}^{\mathsf{H}}{\mathbf{v}}\rvert^2}
		=\frac{{\mathbf v}^{\mathsf H}{\mathbf Q}_{\mathsf b}{\mathbf v}}
		{{\mathbf v}^{\mathsf H}{\mathbf Q}_{\mathsf e}{\mathbf v}}\right),
	\end{equation}
	where ${\mathbf{w}^{\star}}=\sqrt{P}{\mathbf{v}^{\star}}$, and ${\mathbf Q}_{k}={\mathbf I}+{\overline{\gamma}_{k}}{\mathbf h}_{k}{\mathbf h}_{k}^{\mathsf H}\succ{\mathbf{0}}$ for ${k}\in\{{\mathsf b},{\mathsf e}\}$. Problem \eqref{Optimal_Precoding_Trans} is a Rayleigh quotient \cite{zhang2017matrix}, whose optimal solution is given as follows:
	\begin{align}
		{\mathbf{v}^{\star}}=\frac{{\mathbf Q}_{\mathsf e}^{-{1}/{2}}{\mathbf{p}}}{\lVert{\mathbf Q}_{\mathsf e}^{-{1}/{2}}{\mathbf{p}}\rVert}
		=\frac{{\mathbf Q}_{\mathsf e}^{-{1}/{2}}{\mathbf{p}}}{\sqrt{{{\mathbf p}^{\mathsf H}{\mathbf Q}_{\mathsf e}^{-1}{\mathbf p}}}},
	\end{align}
	where ${\mathbf p}$ is the principal eigenvector of the matrix ${\bm\Delta}={\mathbf Q}_{\mathsf e}^{-{1}/{2}}{\mathbf Q}_{\mathsf b}{\mathbf Q}_{\mathsf e}^{-{1}/{2}}\in{\mathbbmss{C}}^{M\times M}$. The resulting optimal objective value in \eqref{Optimal_Precoding_Trans} is given by
	\begin{align}
		\frac{{{\mathbf{v}^{\star}}}^{\mathsf H}{\mathbf Q}_{\mathsf b}{{\mathbf{v}^{\star}}}}
		{{{\mathbf{v}^{\star}}}^{\mathsf H}{\mathbf Q}_{\mathsf e}{{\mathbf{v}^{\star}}}}=\mu_{\bm\Delta},
	\end{align}
	where $\mu_{\bm\Delta}$ is the principal eigenvalue of ${\bm\Delta}$. Consequently, ${\mathbf w}^{\star}$ can be written as follows:
	\begin{align}\label{Optimal_Active_Beamforming_Secrecy_Rate}
		{\mathbf w}^{\star}=\sqrt{P}{{\mathbf v}^{\star}}=\sqrt{P}({{\mathbf p}^{\mathsf H}{\mathbf Q}_{\mathsf e}^{-1}{\mathbf p}})^{-1/2}{{\mathbf Q}_{\mathsf e}^{-{1}/{2}}{\mathbf p}}.
	\end{align} 
	Accordingly, the secrecy capacity is given by 
	\begin{align}
		{\mathsf{C}}=\max\{\log_2{\mu_{\bm\Delta}},0\}.    
	\end{align}
	Leveraging the matrix determinant lemma \cite{horn2012matrix}, we obtain the following theorem.
	\vspace{-5pt}
	\begin{theorem}\label{Theorem_Secrecy_Capacity}
		The secrecy capacity ${\mathsf C}$ is given by
		\begin{align}\label{Basic_Secrecy_Capacity_Exp}
			{\mathsf C}=\log_2\left(1+\frac{\alpha+\sqrt{\alpha^2+4\beta}}{2(1+{\overline{\gamma}}_{\mathsf{e}}G_{\mathsf{e}})}\right),
		\end{align}
		where $\alpha={\overline{\gamma}}_{\mathsf{b}}G_{\mathsf{b}}-{\overline{\gamma}_{\mathsf{e}}}G_{\mathsf{e}}+{\overline{\gamma}_{\mathsf{b}}}{\overline{\gamma}_{\mathsf{e}}}G_{\mathsf{b}}G_{\mathsf{e}}(1-\rho)$, $\beta=(1+{\overline{\gamma}_{\mathsf{e}}}G_{\mathsf{e}}){\overline{\gamma}_{\mathsf{b}}}{\overline{\gamma}_{\mathsf{e}}}G_{\mathsf{b}}G_{\mathsf{e}}(1-\rho)$, $G_k=\lVert{\mathbf{h}}_k\rVert^2$ denotes the channel gain of node $k\in\{{\mathsf{b}},{\mathsf{e}}\}$, and $\rho=\frac{\lvert{\mathbf{h}}_{\mathsf{b}}^{\mathsf{H}}{\mathbf{h}}_{\mathsf{e}}\rvert^2}{\lVert{\mathbf{h}}_{\mathsf{b}}\rVert^2\lVert{\mathbf{h}}_{\mathsf{e}}\rVert^2}\in[0,1]$ denotes the channel correlation factor.
	\end{theorem}
	\vspace{-5pt}
	\begin{IEEEproof}
		Please refer to Appendix \ref{Proof_Theorem_Secrecy_Capacity} for more details.
	\end{IEEEproof}
	By observing \eqref{Basic_Secrecy_Capacity_Exp}, we conclude the following results.
	\vspace{-5pt}
	\begin{corollary}\label{monotone_cor_rho}
		Given $G_{\mathsf{b}}$ and $G_{\mathsf{e}}$, the secrecy capacity ${\mathsf C}$ is monotone decreasing with the correlation factor $\rho\in[0,1]$.
	\end{corollary}
	\vspace{-5pt}
	\begin{IEEEproof}
		Given $G_{\mathsf{b}}$ and $G_{\mathsf{e}}$, only $g_{\mathsf{n}}(\rho)=\alpha+\sqrt{\alpha^2+4\beta}$ in \eqref{Basic_Secrecy_Capacity_Exp} is determined by $\rho$. Furthermore, it holds that
		\begin{equation}
			\frac{{\rm{d}}g_{\mathsf{n}}(\rho)}{{\rm{d}}\rho}=-{\overline{\gamma}_{\mathsf{b}}}{\overline{\gamma}_{\mathsf{e}}}G_{\mathsf{b}}G_{\mathsf{e}}\left(1+\frac{\alpha}{\sqrt{\alpha^2+4\beta}}\right)
			-\frac{1+{\overline{\gamma}_{\mathsf{e}}}G_{\mathsf{e}}}{\sqrt{\alpha^2+4\beta}}.
		\end{equation}
		Since $\rho\in[0,1]$, we have $\beta\geq0$, and it follows that $\lvert{\alpha}/{\sqrt{\alpha^2+4\beta}}\rvert\leq1$ and $1+{\alpha}/{\sqrt{\alpha^2+4\beta}}\geq0$. Thus, we have $\frac{{\rm{d}}g_{\mathsf{n}}(\rho)}{{\rm{d}}\rho}<0$, and the final results follow immediately.
	\end{IEEEproof}
	The above arguments imply that the upper bound of $\mathsf{C}$ is obtained by setting $\rho=0$, which is given as follows.
	\vspace{-5pt}
	\begin{corollary}\label{cor_rho=0}
		When $\rho=0$, the secrecy capacity given in \eqref{Basic_Secrecy_Capacity_Exp} can be written as follows:
		\begin{align}\label{Upper_Bound_Secrecy_Capacity}
			\mathsf{C}=\log _2\left( 1+\overline{\gamma }_{\mathsf{b}}G_{\mathsf{b}} \right). 
		\end{align}
	\end{corollary}
	\vspace{-5pt}
	\begin{IEEEproof}
		When $\rho=0$, we have 
		\begin{subequations}\label{Simplified_Parameters}
			\begin{align}
				\alpha&=\overline{\gamma }_{\mathsf{b}}G_{\mathsf{b}}\left( 1+\overline{\gamma }_{\mathsf{e}}G_{\mathsf{e}} \right) -\overline{\gamma} _{\mathsf{e}}G_{\mathsf{e}} ,\\
				\beta&=(1+{\overline{\gamma}_{\mathsf{e}}}G_{\mathsf{e}}){\overline{\gamma}_{\mathsf{b}}}{\overline{\gamma}_{\mathsf{e}}}G_{\mathsf{b}}G_{\mathsf{e}},\\
				\alpha^2+4\beta&=\left( \overline{\gamma }_{\mathsf{b}}G_{\mathsf{b}}\left( 1+\overline{\gamma }_{\mathsf{e}}G_{\mathsf{e}} \right) +\overline{\gamma} _{\mathsf{e}}G_{\mathsf{e}} \right)^2.
			\end{align}
		\end{subequations}
		Substituting \eqref{Simplified_Parameters} into \eqref{Basic_Secrecy_Capacity_Exp} gives
		\begin{align}
			\mathsf{C}=\log _2\left( 1+\frac{2\overline{\gamma }_{\mathsf{b}}G_{\mathsf{b}}\left( 1+\overline{\gamma }_{\mathsf{e}}G_{\mathsf{e}} \right)}{2(1+\overline{\gamma }_{\mathsf{e}}G_{\mathsf{e}})} \right).
		\end{align}
		The final results follow immediately.
	\end{IEEEproof}
	\vspace{-5pt}
	\begin{remark}
		It is worth noting that \eqref{Upper_Bound_Secrecy_Capacity} is achieved when the received power at Eve is zero, while the received power at Bob is maximized by the maximal-ratio transmission (MRT) beamformer $\mathbf{w}^\star=\sqrt{P}\frac{\mathbf{h}_{\mathsf{b}}}{\| \mathbf{h}_{\mathsf{b}} \|}\triangleq\mathbf{w}_\mathsf{m}$, which serves as an upper bound of the secrecy capacity. This implies that the upper bound of the secrecy capacity is attainable when the channel correlation factor collapses to zero, i.e., $\rho=0$.
	\end{remark}
	\vspace{-5pt}
	\vspace{-5pt}
	\begin{corollary}\label{Collary_Secrecy_Capacity}
		The secrecy capacity collapses to zero if and only if $\rho=1$ and ${\overline{\gamma}_{\mathsf{b}}}G_{\mathsf{b}}\leq{\overline{\gamma}_{\mathsf{e}}}G_{\mathsf{e}}$.
	\end{corollary}
	\vspace{-5pt}
	\begin{IEEEproof}
		When $\mathsf{C} = 0$, it necessitates that $\alpha + \sqrt{\alpha^2 + 4\beta} = 0$. Under this condition, it follows that $\alpha \leq 0$ and $\beta = 0$, which yields ${\overline{\gamma}_{\mathsf{b}}}G_{\mathsf{b}}\leq{\overline{\gamma}_{\mathsf{e}}}G_{\mathsf{e}}$ and $\rho=1$, respectively.
	\end{IEEEproof}
	\vspace{-5pt}
	\begin{remark}\label{remark_insecure}
		The results in Corollary \ref{Collary_Secrecy_Capacity} indicate that secure transmission is unachievable only when the channels of Bob and Eve are parallel and the total signal power gain of Bob is no lager than that of Eve, i.e., ${\overline{\gamma}_{\mathsf{b}}}G_{\mathsf{b}}\leq{\overline{\gamma}_{\mathsf{e}}}G_{\mathsf{e}}$.
	\end{remark}
	\vspace{-5pt}
	In the subsequent subsections, we derive closed-form expressions for the secrecy capacity and conduct asymptotic analyses, considering various channel models ranging from far-field to near-field. Since the capacity becomes zero when Bob and Eve are at the same location, we focus on scenarios where the users are at different locations, i.e., 
	$\mathbf{r}_{\mathsf{b}}\ne\mathbf{r}_{\mathsf{e}}$. For brevity, we assume $\sigma_\mathsf{b}^2=\sigma_\mathsf{e}^2=\sigma^2$ and $\overline{\gamma}_\mathsf{b}=\overline{\gamma}_\mathsf{e}=\overline{\gamma}$. 
	
	\begin{table*}[!t]
		\center
		\resizebox{0.98\textwidth}{!}{
			\begin{tabular}{|c|c|}
				\hline
				\textbf{Model} & \textbf{Secrecy Capacity} \\ \hline
				\textbf{UPW}   & $\log_2\left(1+\frac{{\overline{\gamma}}_{\mathsf{b}}\frac{MA\Psi _\mathsf{b}}{4\pi r_{\mathsf{b}}^{2}}-{\overline{\gamma}_{\mathsf{e}}}\frac{MA\Psi _\mathsf{e}}{4\pi r_{\mathsf{e}}^{2}}+{\overline{\gamma}_{\mathsf{b}}}{\overline{\gamma}_{\mathsf{e}}}
					\frac{M^2A^2\Psi_\mathsf{b}\Psi_\mathsf{e}}{(4\pi)^2r_{\mathsf{b}}^{2}r_{\mathsf{e}}^{2}}(1-\rho_{_{\mathsf{P}}})+
					\sqrt{\left({\overline{\gamma}}_{\mathsf{b}}\frac{MA\Psi _\mathsf{b}}{4\pi r_{\mathsf{b}}^{2}}-{\overline{\gamma}_{\mathsf{e}}}\frac{MA\Psi _\mathsf{e}}{4\pi r_{\mathsf{e}}^{2}}+{\overline{\gamma}_{\mathsf{b}}}{\overline{\gamma}_{\mathsf{e}}}
						\frac{M^2A^2\Psi_\mathsf{b}\Psi_\mathsf{e}}{(4\pi)^2r_{\mathsf{b}}^{2}r_{\mathsf{e}}^{2}}(1-\rho_{_{\mathsf{P}}})\right)^2
						+4\left(1+{\overline{\gamma}}_{\mathsf{e}}\frac{MA\Psi _\mathsf{e}}{4\pi r_{\mathsf{e}}^{2}}\right){\overline{\gamma}_{\mathsf{b}}}{\overline{\gamma}_{\mathsf{e}}}
						\frac{M^2A^2\Psi_\mathsf{b}\Psi_\mathsf{e}}{(4\pi)^2r_{\mathsf{b}}^{2}r_{\mathsf{e}}^{2}}(1-\rho_{_{\mathsf{P}}})}}
				{2\left(1+{\overline{\gamma}}_{\mathsf{e}}\frac{MA\Psi _\mathsf{e}}{4\pi r_{\mathsf{e}}^{2}}\right)}\right)$                \\ \hline
				\textbf{USW}   & $\log_2\left(1+\frac{{\overline{\gamma}}_{\mathsf{b}}\frac{MA\Psi _\mathsf{b}}{4\pi r_{\mathsf{b}}^{2}}-{\overline{\gamma}_{\mathsf{e}}}\frac{MA\Psi _\mathsf{e}}{4\pi r_{\mathsf{e}}^{2}}+{\overline{\gamma}_{\mathsf{b}}}{\overline{\gamma}_{\mathsf{e}}}
					\frac{M^2A^2\Psi_\mathsf{b}\Psi_\mathsf{e}}{(4\pi)^2r_{\mathsf{b}}^{2}r_{\mathsf{e}}^{2}}(1-\rho_{_{\mathsf{S}}})+
					\sqrt{\left({\overline{\gamma}}_{\mathsf{b}}\frac{MA\Psi _\mathsf{b}}{4\pi r_{\mathsf{b}}^{2}}-{\overline{\gamma}_{\mathsf{e}}}\frac{MA\Psi _\mathsf{e}}{4\pi r_{\mathsf{e}}^{2}}+{\overline{\gamma}_{\mathsf{b}}}{\overline{\gamma}_{\mathsf{e}}}
						\frac{M^2A^2\Psi_\mathsf{b}\Psi_\mathsf{e}}{(4\pi)^2r_{\mathsf{b}}^{2}r_{\mathsf{e}}^{2}}(1-\rho_{_{\mathsf{S}}})\right)^2
						+4\left(1+{\overline{\gamma}}_{\mathsf{e}}\frac{MA\Psi _\mathsf{e}}{4\pi r_{\mathsf{e}}^{2}}\right){\overline{\gamma}_{\mathsf{b}}}{\overline{\gamma}_{\mathsf{e}}}
						\frac{M^2A^2\Psi_\mathsf{b}\Psi_\mathsf{e}}{(4\pi)^2r_{\mathsf{b}}^{2}r_{\mathsf{e}}^{2}}(1-\rho_{_{\mathsf{S}}})}}
				{2\left(1+{\overline{\gamma}}_{\mathsf{e}}\frac{MA\Psi _\mathsf{e}}{4\pi r_{\mathsf{e}}^{2}}\right)}\right)$                \\ \hline
				\textbf{NUSW}  & $\log_2\left(1+\frac{{\overline{\gamma}}_{\mathsf{b}}G_{\mathsf{b}}^{\mathsf{N}}-{\overline{\gamma}_{\mathsf{e}}}G_{\mathsf{e}}^{\mathsf{N}}+
					{\overline{\gamma}_{\mathsf{b}}}{\overline{\gamma}_{\mathsf{e}}}G_{\mathsf{b}}^{\mathsf{N}}G_{\mathsf{e}}^{\mathsf{N}}(1-\rho_{_{\mathsf{N}}})
					+\sqrt{\left({\overline{\gamma}}_{\mathsf{b}}G_{\mathsf{b}}^{\mathsf{N}}-{\overline{\gamma}_{\mathsf{e}}}G_{\mathsf{e}}^{\mathsf{N}}+{\overline{\gamma}_{\mathsf{b}}}{\overline{\gamma}_{\mathsf{e}}}
						G_{\mathsf{b}}^{\mathsf{N}}G_{\mathsf{e}}^{\mathsf{N}}(1-\rho_{_{\mathsf{N}}})\right)^2+4(1+{\overline{\gamma}_{\mathsf{e}}}G_{\mathsf{e}}^{\mathsf{N}}){\overline{\gamma}_{\mathsf{b}}}{\overline{\gamma}_{\mathsf{e}}}
						G_{\mathsf{b}}^{\mathsf{N}}G_{\mathsf{e}}^{\mathsf{N}}(1-\rho_{_{\mathsf{N}}})}}{2(1+{\overline{\gamma}}_{\mathsf{e}}G_{\mathsf{e}}^{\mathsf{N}})}\right)$                \\ \hline
		\end{tabular}}
		\caption{Secrecy channel capacity under different channel models.}
		\vspace{-5pt}
		\label{table0}
	\end{table*}
	
	\subsection{Far-Field and Near-Field Secrecy Capacity}
	\subsubsection{UPW Model}
	When the users are located in the far-field region where the UPW channel model is applicable, a closed-form expression for $\mathsf{C}$ is given as follows.
	\vspace{-5pt}
	\begin{theorem}\label{UPW_the}
		Under the UPW model, the channel gains are given by $G_k=\frac{MA\Psi _k}{4\pi r_{k}^{2}}\triangleq G_k^{\mathsf{P}}$ for $k\in\{\mathsf{b},\mathsf{e}\}$, and the channel correlation factor is given by $\rho=\rho _{_{\mathsf{P}}}$, where
		\begin{align}\label{rho_upw}
			\rho _{_{\mathsf{P}}} \!=\!\left\{\! \!\!\begin{array}{ll}
				1&	\!\!	\left( \theta _\mathsf{b},\phi _\mathsf{b} \right)\! =\!\left( \theta _\mathsf{e},\phi _\mathsf{e} \right)\\
				\frac{1-\cos \left( M_x\Xi _{\Phi} \right)}{M^2\left( 1-\cos \Xi _{\Phi} \right)}&	\!\!	\Phi _\mathsf{b}\!\ne\! \Phi _\mathsf{e},\Omega _\mathsf{b}\!=\!\Omega _\mathsf{e}\\
				\frac{1-\cos \left( M_z\Xi _{\Omega} \right)}{M^2\left( 1-\cos \Xi _{\Omega} \right)}&	\!\!	\Phi _\mathsf{b}\!=\!\Phi _\mathsf{e},\Omega _\mathsf{b}\!\ne\! \Omega _\mathsf{e}\\
				\frac{4\left( 1-\cos \left( M_x\Xi _{\Phi} \right) \right) \left( 1-\cos \left( M_z\Xi _{\Omega} \right) \right)}{M^2\left( 1-\cos \Xi _{\Phi} \right) \left( 1-\cos \Xi _{\Omega} \right)}&	\!\!	\text{else}\\
			\end{array} \right.\!\!\!, 
		\end{align}
		and where $\Xi _{\Phi}=\frac{2\pi}{\lambda}d\left( \Phi _\mathsf{b}-\Phi _\mathsf{e} \right)$, and $\Xi _{\Omega}=\frac{2\pi}{\lambda}d\left( \Omega _\mathsf{b}-\Omega _\mathsf{e} \right)$. The secrecy capacity for FFC is shown in Table \ref{table0}.
	\end{theorem}
	\vspace{-5pt}
	\begin{IEEEproof}
		The results can be derived from \eqref{channel_FF} using the sum of the geometric series and trigonometric identities.
	\end{IEEEproof}
	We next consider a challenging secure communication scenario, where the Eve is located in the same direction of the Bob, i.e., $\left( \theta _\mathsf{b},\phi _\mathsf{b} \right) =\left( \theta _\mathsf{e},\phi _\mathsf{e} \right)$. According to \eqref{rho_upw}, it holds that $\rho _{_{\mathsf{P}}}=1$ for $\left( \theta _\mathsf{b},\phi _\mathsf{b} \right) =\left( \theta _\mathsf{e},\phi _\mathsf{e} \right)$, and thus the secrecy capacity is calculated as follows.
	\vspace{-5pt}
	\begin{corollary}\label{cor_upw}
		When $\left( \theta _\mathsf{b},\phi _\mathsf{b} \right) =\left( \theta _\mathsf{e},\phi _\mathsf{e} \right)$, the secrecy capacity under the UPW model can be written as
		\begin{align}\label{upw_same}
			\mathsf{C}=\begin{cases}
				\log _2\Big( \frac{1+\overline{\gamma}G_{\mathsf{b}}^{\mathsf{P}}}{1+\overline{\gamma}G_{\mathsf{e}}^{\mathsf{P}}} \Big)&		r_{\mathsf{b}}<r_{\mathsf{e}}\\
				0&		r_{\mathsf{b}}\ge r_{\mathsf{e}}\\
			\end{cases}.
		\end{align}
	\end{corollary}
	\vspace{-5pt}
	\begin{remark}\label{rem_upw_C}
		The above arguments imply that under the FFC model, secure transmission becomes unachievable when both users are located in the same direction, and Bob is at a distance from the BS that is greater than or equal to Eve's.
	\end{remark}
	\vspace{-5pt}
	To gain further insights into the limits of secrecy performance, we next analyze the secrecy capacity under two asymptotic conditions: \romannumeral1) where the number of BS antennas approaches infinity, i.e., $M_x,M_z\rightarrow\infty $ (i.e., $M\rightarrow\infty$), and \romannumeral2) where the transmit power approaches infinity, i.e., 
	$P\rightarrow\infty$.
	
	We first consider the case of $M\rightarrow\infty$.
	\vspace{-5pt}
	\begin{corollary}\label{cor_upw_M}
		As $M\rightarrow\infty$, the secrecy capacity for the UPW model satisfies
			\begin{align}
				\lim_{M\rightarrow\infty}\mathsf{C} \begin{cases}
					=\max \{ 2\log _2\frac{r_{\mathsf{e}}}{r_{\mathsf{b}}}, 0 \}&		\left( \theta _{\mathsf{b}},\phi _{\mathsf{b}} \right) \!=\!\left( \theta _{\mathsf{e}},\phi _{\mathsf{e}} \right)\\
					\simeq\mathcal{O} (\log _2M)&		\left( \theta _{\mathsf{b}},\phi _{\mathsf{b}} \right) \!\ne\! \left( \theta _{\mathsf{e}},\phi _{\mathsf{e}} \right)\\
				\end{cases}.
		\end{align}  
	\end{corollary}
	\begin{IEEEproof}
		For $\left( \theta _{\mathsf{b}},\phi _{\mathsf{b}} \right) =\left( \theta _{\mathsf{e}},\phi _{\mathsf{e}} \right)$, it follows from \eqref{upw_same} that $\lim_{M\rightarrow\infty}\mathsf{C}=\max \{ 2\log _2\frac{r_{\mathsf{e}}}{r_{\mathsf{b}}}, 0 \}$. For $\left( \theta _{\mathsf{b}},\phi _{\mathsf{b}} \right) \ne \left( \theta _{\mathsf{e}},\phi _{\mathsf{e}} \right)$, we have $\lim_{M\rightarrow\infty}\rho _{_{\mathsf{P}}}=0$, which, together with Corollary \ref{cor_rho=0}, yields
		$\mathsf{C}=\log _2\left( 1+\overline{\gamma }G_{\mathsf{b}}^{\mathsf{P}} \right)\simeq\mathcal{O} (\log _2M)$.  
	\end{IEEEproof}
	\vspace{-2pt}
	\begin{remark}\label{rem_upw_M}
		The results of Corollary \ref{cor_upw_M} suggest that the secrecy capacity under the UPW model may increase unboundedly with the number of transmit antennas, theoretically reaching any desired level. However, this contradicts the principles of energy conservation.
	\end{remark}

	When $P\rightarrow\infty$, it has ${\overline{\gamma}}\rightarrow\infty$, which yields Corollary \ref{cor_upw_p}.
	\begin{corollary}\label{cor_upw_p}
		When ${\overline{\gamma}}\rightarrow\infty$, the secrecy capacity satisfies
		\begin{align}
			\mathsf{C}\simeq \begin{cases}
				\max \left\{ 2\log _2\frac{r_{\mathsf{e}}}{r_{\mathsf{b}}},0 \right\}&		\left( \theta _{\mathsf{b}},\phi _{\mathsf{b}} \right) =\left( \theta _{\mathsf{e}},\phi _{\mathsf{e}} \right)\\
				\log _2\left( \overline{\gamma }G_{\mathsf{b}}^{\mathsf{P}}(1-\rho_{_{\mathsf{P}}} ) \right)&		\left( \theta _{\mathsf{b}},\phi _{\mathsf{b}} \right) \ne \left( \theta _{\mathsf{e}},\phi _{\mathsf{e}} \right)\\
			\end{cases}.
		\end{align}
	\end{corollary}
	\begin{IEEEproof}
		Please refer to Appendix~\ref{proof_cor_p} for more details.
	\end{IEEEproof}
	
	\begin{remark}
		The results of Corollary~\ref{cor_upw_p} suggest that under the UPW channel model, the high-SNR slope of the secrecy capacity is zero when Bob and Eve are in the same direction, indicating that increasing the transmit power does not improve the secrecy capacity. However, when Bob and Eve are in different directions, the high-SNR slope is one, meaning that the secrecy capacity increases linearly with the logarithm of the transmit power.  
	\end{remark}
	
	\begin{figure}[!t]
		\centering
		\includegraphics[height=0.27\textwidth]{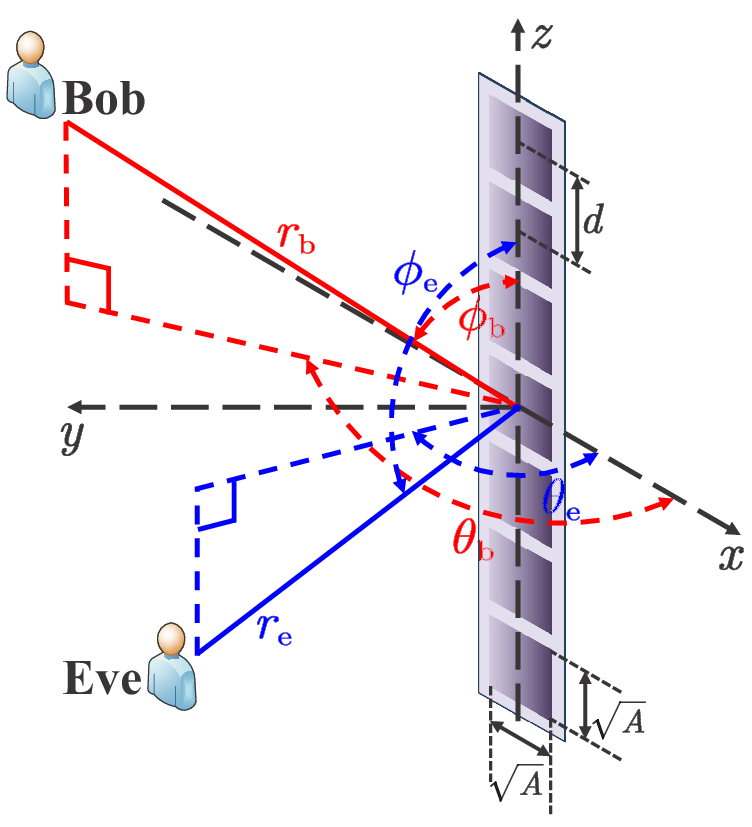}
		\caption{Illustration of the ULA.}
		\label{ula}
		\vspace{-5pt}
	\end{figure}
	
	Next, we consider a special case where the BS is equipped with a uniform linear array (ULA). In this case, we have $M_x=1$ and $M_z=M$, as illustrated in Fig.~\ref{ula}.
		\begin{corollary}
			With a ULA, the channel gains under the UPW model are given by $G_k=\frac{MA\Psi _k}{4\pi r_{k}^{2}}$ for $k\in\{\mathsf{b},\mathsf{e}\}$, and the channel correlation factor is given by
			\begin{align}
				\rho =\begin{cases}
					1&		\phi _{\mathsf{b}}=\phi _{\mathsf{e}}\\
					\frac{1-\cos \left( M\varDelta _{\Omega} \right)}{M^2\left( 1-\cos \varDelta _{\Omega} \right)}&		\phi _{\mathsf{b}}\ne \phi _{\mathsf{e}}\\
				\end{cases}
			\end{align}
		\end{corollary}
		\vspace{-5pt}
		\begin{IEEEproof}
			Similar to the proof of Theorem~\ref{UPW_the}.
		\end{IEEEproof}
		Having derived the expressions of the channel gains and the correlation factor for the ULA case, we can readily obtain the secrecy capacity and perform the similar analysis as in the UPA scenario, which are omitted due to space limitation.
		\subsubsection{USW Model}
		We now turn to the NFC, commencing with the scenario where the users are located between the Rayleigh and uniform-power distances. In this case, the USW model is applicable, which yields the following theorem.
		\vspace{-5pt}
		\begin{theorem}\label{USW_the}
			Under the USW channel model, the channel gains are given by $G_k=\frac{MA\Psi _k}{4\pi r_{k}^{2}}\triangleq G_k^{\mathsf{S}}$ for $k\in\{\mathsf{b},\mathsf{e}\}$, and the channel correlation factor satisfies $\rho =\frac{\delta _x \delta _z}{M^2} \triangleq  \rho _{_{\mathsf{S}}}$, where 
			\begin{align}\label{delta}
				\delta_i \!\approx\!\left\{\! \!\!\begin{array}{ll}
					\left| \frac{\sin \left( M_ib_i/2 \right)}{\sin \left( b_i/2 \right)} \right|^2&		\!\!\!a_i\!=\!0, b_i\!\ne\! 0\\
					\frac{\pi}{\left| a_i \right|}\left| \mathrm{erf}\left( \frac{M_i}{2}\sqrt{\left| a_i \right|}\mathrm{e}^{\mathrm{j}\frac{\pi}{4}} \right) \right|^2&		\!\!\!a_i\!\ne \!0, b_i\!=\!0\\
					\!\frac{\pi}{4\left| a_i \right|}\!\!\left| \mathrm{erf}\Big(\! \frac{ a_iM_i\!-b_i }{2\sqrt{\left| a_i \right|}}\mathrm{e}^{\mathrm{j}\frac{\pi}{4}} \!\!\Big) \!\!+\!\mathrm{erf}\Big( \!\frac{ a_iM_i\!+b_i }{2\sqrt{\left| a_i \right|}}\mathrm{e}^{\mathrm{j}\frac{\pi}{4}}\!\! \Big) \!\right|^2&		\!\!\!a_i\!\ne\! 0,b_i\!\ne\! 0\\
				\end{array}\right.\!\!     
			\end{align}
			for $i\in \left\{ x,z \right\}$, and where $b_x=\frac{2\pi d}{\lambda}\left( \Phi _{\mathsf{b}}-\Phi _{\mathsf{e}} \right) $, $b_z=\frac{2\pi d}{\lambda}\left( \Omega _{\mathsf{b}}-\Omega _{\mathsf{e}} \right) $, $a_x=\frac{\pi d^2}{\lambda}\left( \frac{1-\Phi _{\mathsf{b}}^{2}}{r_{\mathsf{b}}}-\frac{1-\Phi _{\mathsf{e}}^{2}}{r_{\mathsf{e}}} \right)  $, $a_z=\frac{\pi d^2}{\lambda}\left( \frac{1-\Omega _{\mathsf{b}}^{2}}{r_{\mathsf{b}}}-\frac{1-\Omega _{\mathsf{e}}^{2}}{r_{\mathsf{e}}} \right) $, and $\mathrm{erf}\left( x \right) =\frac{2}{\sqrt{\pi}}\int_0^x{\mathrm{e}^{-t^2}\mathrm{d}t}$ is the error function \cite[Eq. (8.250.1)]{integral}. The expression for the secrecy capacity is shown in Table \ref{table0}. 
		\end{theorem}
		\vspace{-5pt}
		\begin{IEEEproof}
			Please refer to Appendix~\ref{proof_usw} for more details.
		\end{IEEEproof}
		Similar to the analysis of the UPW model, we also examine the asymptotic secrecy capacity for the USW model in the following discussions.
		\vspace{-5pt}
		\begin{corollary}\label{cor_usw_M}
			When $M\rightarrow\infty$, the secrecy capacity for the USW model satisfies
		\begin{equation}
			\lim\nolimits_{M\rightarrow\infty}\mathsf{C}\simeq {\mathcal{O}}(\log_2{M}).
		\end{equation}   
		\end{corollary}
		\vspace{-5pt}
		\begin{IEEEproof}
			Similar to the proof of Corollary~\ref{cor_upw_M}. 
		\end{IEEEproof}
		\vspace{-3pt}
		\begin{remark}\label{rem_usw_M}
			The results of Corollary~\ref{cor_usw_M} suggest that the secrecy capacity under the USW model may increase indefinitely with the number of transmit antennas, which contradicts the laws of energy conservation.
		\end{remark}
		\vspace{-3pt}
		Then we consider the case of $P\rightarrow\infty$.
		\vspace{-5pt}
		\begin{corollary}\label{cor_usw_p}
			When ${\overline{\gamma}}\rightarrow\infty$, the secrecy capacity satisfies
			\begin{align}
				\mathsf{C}\simeq
				\log _2(\overline{\gamma }G_{\mathsf{b}}^{\mathsf{S}}(1-\rho_{_{\mathsf{S}}} ) ).
			\end{align}
		\end{corollary}
		\vspace{-5pt}
		\begin{IEEEproof}
			Similar to the proof of Corollary~\ref{cor_upw_p}.    
		\end{IEEEproof}
		\vspace{-5pt}
		\begin{remark}
			The results of Corollary~\ref{cor_usw_p} suggest that the high-SNR slope of the capacity under the USW model is one.   
		\end{remark}
		We then consider the ULA case under the USW model.
			\begin{corollary}
				When a ULA is employed, under the USW channel model, the channel gains are given by $G_k=\frac{MA\Psi _k}{4\pi r_{k}^{2}}$ for $k\in\{\mathsf{b},\mathsf{e}\}$, and the channel correlation factor satisfies $\rho =\frac{ \delta _z}{M^2}$.
			\end{corollary}
			\vspace{-3pt}
			\begin{IEEEproof}
				Similar to the proof of Theorem~\ref{USW_the}.
		\end{IEEEproof}
		
		\subsubsection{NUSW Model}
		Next, we investigate the NUSW model for the users positioned within the uniform-power distance.
		\vspace{-3pt}
		\begin{theorem}\label{the_nusw}
			Under the NUSW model, the resulting channel gain of node $k\in\{{\mathsf{b}},{\mathsf{e}}\}$ can be expressed as follows:
			\begin{align}\label{Expression_Channel_Gain}
				G_k\!=\!\frac{A}{4\pi d^2}\sum_{x\in{\mathcal{X}}_k}\!\sum_{z\in{\mathcal{Z}}_k}\!\arctan\!\bigg(\frac{xz\Psi_k^{-1}}{\sqrt{\Psi_k^2\!+\!x^2\!+\!z^2}}\bigg)\!\triangleq G_k^{\mathsf{N}},
			\end{align}
			where ${\mathcal{X}}_k=\{\frac{M_x}{2}\varepsilon_k\pm\Phi_k\}$ and ${\mathcal{Z}}_k=\{\frac{M_z}{2}\varepsilon_k\pm\Psi_k\}$. As for the channel correlation factor, it can be evaluated as follows:
			\begin{align}\label{Expression_Squared_Correlation_Coefficient}
				&\rho =\frac{M^2A^2\Psi _{\mathsf{b}}\Psi _{\mathsf{e}}\pi ^2}{16G_{\mathsf{b}}^{\mathsf{N}}G_{\mathsf{e}}^{\mathsf{N}}r_{\mathsf{b}}^{2}r_{\mathsf{e}}^{2}T^4}\left| \sum\nolimits_{t=1}^T\sum\nolimits_{t^{\prime}=1}^T{\sqrt{\left( 1-\zeta _{t}^{2} \right) \left( 1-\zeta _{t^{\prime}}^{2} \right)}}\right.\nonumber\\
				&\left.\times g_1\!\left( M_x\varepsilon _{\mathsf{b}}\zeta _t,M_z\varepsilon _{\mathsf{b}}\zeta _{t^{\prime}} \right) g_2\!\left( M_x\varepsilon _{\mathsf{b}}\zeta _t,M_z\varepsilon _{\mathsf{b}}\zeta _{t^{\prime}} \right) \right|^2\triangleq  \rho _{_{\mathsf{N}}},
			\end{align}
			where $T$ is a complexity-vs-accuracy tradeoff parameter, $\zeta _t=\cos \left( \frac{2t-1}{2T}\pi \right) $ for $t\in\left\{ 1,\ldots,T \right\} $, and 
			\begin{subequations}
				\begin{align}
					g_1\left( x,z \right) &\triangleq\frac{\mathrm{e}^{\mathrm{j}\frac{2\pi}{\lambda}r_{\mathsf{b}}\left( x^2+z^2-2\Phi _{\mathsf{b}}x-2\Omega _{\mathsf{b}}z+1 \right) ^{\frac{1}{2}}}}{\left( x^2+z^2-2\Phi _{\mathsf{b}}x-2\Omega _{\mathsf{b}}z+1 \right) ^{\frac{3}{2}}}   ,\\
					g_2\left( x,z \right) &\triangleq\frac{\mathrm{e}^{-\mathrm{j}\frac{2\pi}{\lambda}r_{\mathsf{e}}\left( \tau ^2x^2+\tau ^2z^2-2\tau \Phi _{\mathsf{e}}x-2\tau \Omega _{\mathsf{e}}z+1 \right) ^{\frac{1}{2}}}}{\left( \tau ^2x^2+\tau ^2z^2-2\tau \Phi _{\mathsf{e}}x-2\tau \Omega _{\mathsf{e}}z+1 \right) ^{\frac{3}{2}}}
				\end{align}
			\end{subequations}
			with $\tau =\frac{r_{\mathsf{b}}}{r_{\mathsf{e}}}$. The secrecy capacity is shown in Table \ref{table0}.
		\end{theorem}
		\vspace{-5pt}
		\begin{IEEEproof}
			Please refer to Appendix \ref{proof_nusw} for more details.
		\end{IEEEproof}
		Due to the linearly varying phase shifts in the UPW model, the channels for Bob and Eve become parallel when they are located in the same direction, resulting in $\rho=1$ and thus $\mathsf{C}=0$ for $r_{\mathsf{b}}\ge r_{\mathsf{e}}$. In contrast, the nonlinear varying phase shifts in the spherical-wave models, i.e., USW and NUSW, prevent the channels for Bob and Eve from aligning perfectly unless they are at the exact same location. According to Corollary~\ref{Collary_Secrecy_Capacity}, we draw the following conclusion.
		\vspace{-5pt}
		\begin{remark}\label{rem_usw_C}
			Unlike the UPW model, the secrecy capacity under the USW and NUSW models does not collapse to zero when Bob and Eve are positioned in the same direction but at different locations.
		\end{remark}
		
		These arguments suggest that leveraging the near-field effect can effectively enhance the system's secrecy performance. The superiority of NFC is essentially attributed to spherical-wave EM propagation, which allows energy to be focused on Bob's specific location ${\mathbf{r}}_{\mathsf{b}}$ rather than merely in his direction $(\theta_\mathsf{b},\phi_\mathsf{b})$. This capability of \emph{beamfocusing} significantly improves secrecy performance in contrast to FFC.
		
		\begin{table}[!t]
			\center
			\scalebox{0.8}{
					\begin{tabular}{|c|c|c|c|}\hline
						{\multirow{2}{*}{Model}} & {\multirow{2}{*}{Energy conservation}} & \multicolumn{2}{c|}{When $\left( \theta _\mathsf{b},\phi _\mathsf{b} \right)\! =\!\left( \theta _\mathsf{e},\phi _\mathsf{e} \right)$}  \\ \cline{3-4}
						& &$\mathsf{C}$  & High-SNR slope \\ \hline
						UPW & $\times$ & $\begin{array}{c}
							\text{If} \ r_{\mathsf{b}}\ge r_{\mathsf{e}}, \ =0\\
							\text{If} \ r_{\mathsf{b}}<r_{\mathsf{e}},\  >0\\
						\end{array}$  & $0$  \\ \hline
						USW & $\times$ & $>0$  & $1$  \\ \hline
						NUSW & $\checkmark$ & $>0$  & $1$  \\ \hline
			\end{tabular}}
			\caption{Comparison between different channel models.}
			\vspace{-5pt}
			\label{table1}
		\end{table}
		
		We next analyze the asymptotic secrecy capacity by setting $M_x,M_z\rightarrow\infty$ and $P\rightarrow\infty$, respectively.
		\vspace{-5pt}
		\begin{corollary}\label{cor_nusw_M}
			When $M_x,M_z\rightarrow\infty$, the secrecy capacity for the NUSW model satisfies
			\begin{align}
				\lim_{M_x,M_z\rightarrow\infty}\mathsf{C}\approx\log _2\left( 1+\frac{\overline{\gamma }A}{2d^2} \right) .
			\end{align}    
		\end{corollary}
		\vspace{-5pt}
		\begin{IEEEproof}
			It follows from the numerical results in \cite{NFISAC_performance} that $\lim_{M_x,M_z\rightarrow \infty}\rho _{_{\mathsf{N}}}\ll 1$, which yields $\mathsf{C}\approx
			\log _2\left( 1+\overline{\gamma }G_{\mathsf{b}}^{\mathsf{N}} \right)$. Furthermore, we have 
			\begin{equation}
				\!\lim_{M_x,M_z\rightarrow \infty}\!G_{k}^{\mathsf{N}}\!=\!\frac{4A\!\lim\limits_{x,z\rightarrow \infty}\!\arctan \!\Big( \frac{xz\Psi _{k}^{-1}}{\sqrt{\Psi _{k}^{2}\!+x^2\!+z^2}} \Big)}{4\pi d^2} \!=\frac{A}{2d^2}.
			\end{equation}
			The final results follow immediately. 
		\end{IEEEproof}
		\vspace{-5pt}
		\begin{remark}\label{rem_nusw_M}
			The results in Corollary~\ref{cor_nusw_M} indicate that, unlike the UPW and USW models, the secrecy capacity under the NUSW model converges to a finite upper bound as the number of transmit antennas increases, thereby adhering to the principle of energy conservation.
		\end{remark}
		\vspace{-5pt}
		Then we consider the case of $P\rightarrow\infty$, i.e., ${\overline{\gamma}}\rightarrow\infty$.
		\vspace{-5pt}
		\begin{corollary}\label{cor_nusw_p}
			When ${\overline{\gamma}}\rightarrow\infty$, the secrecy capacity satisfies
			\begin{align}
				\mathsf{C}\simeq
				\log _2\left( \overline{\gamma }G_{\mathsf{b}}^{\mathsf{N}}(1-\rho_{_{\mathsf{N}}} ) \right).
			\end{align}
		\end{corollary}
		\vspace{-5pt}
		\begin{IEEEproof}
			Similar to the proof of Corollary~\ref{cor_upw_p}.    
		\end{IEEEproof}
		\vspace{-5pt}
		\begin{remark}
			The results in Corollary~\ref{cor_nusw_p} suggest that the high-SNR slope under the NUSW model is one.  
		\end{remark}
		
		We next investigate the ULA case in the following corollary.
			\begin{corollary}\label{cor_ula}
				Under the NUSW model, the channel gain of node $k\in\{{\mathsf{b}},{\mathsf{e}}\}$ is given by 
				\begin{equation}\label{ula_channelgain}
					\begin{split}
						G_k=&\frac{A\varepsilon _k\sin \phi _k}{4\pi d^2\sin \theta _k}\bigg( \frac{M\varepsilon _k-2\cos \theta _k}{\sqrt{M^2\varepsilon _{k}^{2}-4M\cos \theta _k\varepsilon _k+4}}\\
						&+\frac{M\varepsilon _k+2\cos \theta _k}{\sqrt{M^2\varepsilon _{k}^{2}+4M\cos \theta _k\varepsilon _k+4}} \bigg).
					\end{split}
				\end{equation}
				The channel correlation factor for the ULA case can be written as follows:
				\begin{equation}\label{ula_correlation}
					\begin{split}
						&\rho =\frac{M^2A^2\Psi _{\mathsf{b}}\Psi _{\mathsf{e}}\pi ^2}{16G_{\mathsf{b}}^{\mathsf{N}}G_{\mathsf{e}}^{\mathsf{N}}r_{\mathsf{b}}^{2}r_{\mathsf{e}}^{2}T^4}\\
						&~~\times\left| \sum\nolimits_{t=1}^T{\sqrt{\left( 1-\zeta _{t}^{2} \right) }} g'_1\left( M\varepsilon _{\mathsf{b}}\zeta _{t} \right) g'_2\left( M\varepsilon _{\mathsf{b}}\zeta _{t} \right) \right|^2,
					\end{split}	
				\end{equation}
				where $g'_i(z)\triangleq g_i(0,z)$ for $i=1,2$.
			\end{corollary}
			\vspace{-5pt}
			\begin{IEEEproof}
				Please refer to Appendix~\ref{proof_ula} for more details.
		\end{IEEEproof}
		
		For comparison, we summarize key outcomes from the above analysis across different channel models in Table \ref{table1}, leading to the following observations. 
		\vspace{-5pt}
		\begin{remark}
			The NUSW model provides a more realistic approach for energy considerations compared to other models. This underscores the critical importance of precise channel modeling in the near-field region to accurately capture the physical behaviors of EM waves.    
		\end{remark}
		\vspace{-5pt}
		\begin{remark}
			When Eve is aligned in the same direction as Bob, the reduced channel correlation in NFC provides additional resolution in the distance domain, enabling NFC to achieve better secrecy performance compared to FFC.
		\end{remark}
		
		\subsection{Depth of Insecurity}
		As discussed in Remark \ref{rem_usw_C}, in NFC, secure transmission can be ensured even when users are located in the same direction due to the application of \emph{beamfocusing}. Additionally, as proven in Corollary \ref{monotone_cor_rho}, a low channel correlation corresponds to a higher secrecy rate, which can be achieved by equipping the BS with a sufficiently large number of antennas. Corollary \ref{cor_rho=0} further elaborates that when $\rho=0$, the upper bound of the secrecy capacity can be achieved using the MRT beamformer. However, achieving $\rho=0$ is theoretically feasible only with an infinitely large number of antennas, which is impractical. 
		
		Given this impracticality, it becomes crucial to examine the gap between the practical optimal beamformer ${\mathbf{w}^{\star}}$ and the MRT beamformer $\mathbf{w}_\mathsf{m}$. According to \eqref{Optimal_Active_Beamforming_Secrecy_Rate}, a closed-form expression for the optimal beamformer is unavailable, making this comparison particularly challenging. As a compromise, we focus on the optimal beamformer in the asymptotic case where the transmit power $P\rightarrow \infty$, or equivalently, where both $\overline{\gamma}_\mathsf{b}$ and $\overline{\gamma}_\mathsf{e}$ approach infinity. 
		
		When $\overline{\gamma}_\mathsf{b}\rightarrow \infty$, the matrix $\mathbf{\Delta}$ can be approximated by
		\begin{align}
			\mathbf{\Delta }=\mathbf{Q}_{\mathsf{e}}^{-1/2}\mathbf{Q}_{\mathsf{b}}\mathbf{Q}_{\mathsf{e}}^{-1/2}\simeq \overline{\gamma} _{\mathsf{b}}\mathbf{Q}_{\mathsf{e}}^{-1/2}\mathbf{h}_{\mathsf{b}}\mathbf{h}_{\mathsf{b}}^{\mathsf{H}}\mathbf{Q}_{\mathsf{e}}^{-1/2},
		\end{align}
		which implies that its principal eigenvector $\mathbf{p}$ is asymptotically parallel to $\mathbf{Q}_{\mathsf{e}}^{-1/2}\mathbf{h}_{\mathsf{b}}$. Hence, the optimal beamformer is parallel to $\mathbf{Q}_{\mathsf{e}}^{-1}\mathbf{h}_{\mathsf{b}}$, which can be calculated as follows:
			\begin{align}
				\!\!\mathbf{Q}_{\mathsf{e}}^{-1}\mathbf{h}_{\mathsf{b}}&=\left( \mathbf{I}+\overline{\gamma}_{\mathsf{e}}\mathbf{h}_{\mathsf{e}}\mathbf{h}_{\mathsf{e}}^{\mathsf{H}} \right) ^{-1}\mathbf{h}_{\mathsf{b}}\nonumber\\
				&\overset{\left( a \right)}{=}\bigg( \mathbf{I}\!-\frac{\overline{\gamma} _{\mathsf{e}}\mathbf{h}_{\mathsf{e}}\mathbf{h}_{\mathsf{e}}^{\mathsf{H}}}{1+\overline{\gamma} _{\mathsf{e}}\!\left\| \mathbf{h}_{\mathsf{e}} \right\| ^2} \bigg) \mathbf{h}_{\mathsf{b}}\notag\\
				&\overset{\left( b \right)}{\approx} \bigg( \mathbf{I}-\frac{\mathbf{h}_{\mathsf{e}}\mathbf{h}_{\mathsf{e}}^{\mathsf{H}}}{\left\| \mathbf{h}_{\mathsf{e}} \right\| ^2} \bigg) \mathbf{h}_{\mathsf{b}},
			\end{align}
		where the step $(a)$ follows from the Woodbury matrix identity, and the step $(b)$ follows from the assumption of $\overline{\gamma}_\mathsf{e}\rightarrow \infty$. As a result, the optimal beamformer is given by
			\begin{align}
				\mathbf{w}_{\infty}^{\star}&=\sqrt{P}\eta \left( \mathbf{I}-\frac{\mathbf{h}_{\mathsf{e}}\mathbf{h}_{\mathsf{e}}^{\mathsf{H}}}{\| \mathbf{h}_{\mathsf{e}} \| ^2} \right) \mathbf{h}_{\mathsf{b}}=\sqrt{P}\eta \left( \mathbf{h}_{\mathsf{b}}-\frac{\mathbf{h}_{\mathsf{e}}^{\mathsf{H}}\mathbf{h}_{\mathsf{b}}}{\|\mathbf{h}_{\mathsf{e}}\| ^2}\mathbf{h}_{\mathsf{e}}\right),  \label{Asymptotic_Optimal_Beamforming}  
			\end{align}
		where $\eta =\| ( \mathbf{I}-\frac{\mathbf{h}_{\mathsf{e}}\mathbf{h}_{\mathsf{e}}^{\mathsf{H}}}{\| \mathbf{h}_{\mathsf{e}} \| ^2} ) \mathbf{h}_{\mathsf{b}} \| ^{-1}$ is for normalization. Notably, $\mathbf{w}_{\infty}^{\star}$ is orthogonal to $\mathbf{h}_{\mathsf{e}}$, i.e.,
		\begin{align}
			\mathbf{h}_{\mathsf{e}}^{\mathsf{H}}\mathbf{w}_{\infty}^{\star}=\sqrt{P}\eta \left( \mathbf{h}_{\mathsf{e}}^{\mathsf{H}}\mathbf{h}_{\mathsf{b}}-
			\frac{\| \mathbf{h}_{\mathsf{e}} \| ^2}{\| \mathbf{h}_{\mathsf{e}} \| ^2}\mathbf{h}_{\mathsf{e}}^{\mathsf{H}}\mathbf{h}_{\mathsf{b}} \right) =0.
		\end{align}
		\vspace{-5pt}
		\begin{remark}
			The results in \eqref{Asymptotic_Optimal_Beamforming} suggest that in the high-SNR regime, the optimal beamformer aligns with $\mathbf{h}_{\mathsf{b}}$ minus its projection on $\mathbf{h}_{\mathsf{e}}$, which represents the vector most aligned with $\mathbf{h}_{\mathsf{b}}$ within the null space of $\mathbf{h}_{\mathsf{e}}$. 
		\end{remark}
		\vspace{-5pt}
		\vspace{-5pt}
		\begin{remark}
			The above arguments imply that in the high-SNR regime, the secrecy capacity-achieving vector tends to be orthogonal to Eve’s channel. Essentially, when orthogonality to $\mathbf{h}_{\mathsf{e}}$ is achieved, i.e., when Eve's channel gain is minimized, the optimal beamformer $\mathbf{w}_{\infty}^{\star}$ aligns as closely as possible with the direction of $\mathbf{h}_{\mathsf{b}}$. As the number of antennas $M$ approaches infinity, the beamformer coincides with $\mathbf{h}_{\mathsf{b}}$.
		\end{remark}
		\vspace{-5pt}
		Therefore, we can use the angle between $\mathbf{w}_{\infty}^{\star}$ and $\mathbf{w}_\mathsf{m}=\sqrt{P}\frac{\mathbf{h}_{\mathsf{b}}}{\| \mathbf{h}_{\mathsf{b}} \|}$, denoted as $\psi\in[0,\pi]$, to numerically evaluate the gap between them. The cosine of $\psi$, which decreases monotonically as $\psi$ increases, is given in Lemma \ref{lem_angle}.
		\vspace{-5pt}
		\begin{lemma}\label{lem_angle}
			The cosine of $\psi$ is calculated as follows:
			\begin{align}
				\cos \psi=1-\rho.
			\end{align}
		\end{lemma}
		\vspace{-5pt}
		\begin{IEEEproof}
			By definition, we have $\cos\psi=\frac{| \mathbf{w}_{\mathsf{m}}^{\mathsf{H}}\mathbf{w}_{\infty}^{\star} |^2}{\| \mathbf{w}_{\mathsf{m}} \| ^2\| \mathbf{w}_{\infty}^{\star} \| ^2}$, which yields $\cos \psi=\frac{\big| \frac{\mathbf{h}_{\mathsf{b}}^{\mathsf{H}}}{\| \mathbf{h}_{\mathsf{b}} \|}\big( \mathbf{h}_{\mathsf{b}}-\frac{\mathbf{h}_{\mathsf{e}}^{\mathsf{H}}\mathbf{h}_{\mathsf{b}}}{\| \mathbf{h}_{\mathsf{e}} \| ^2}\mathbf{h}_{\mathsf{e}} \big) \big|^2}{\mathbf{h}_{\mathsf{b}}^{\mathsf{H}}\big( \mathbf{I}-\frac{\mathbf{h}_{\mathsf{e}}\mathbf{h}_{\mathsf{e}}^{\mathsf{H}}}{\| \mathbf{h}_{\mathsf{e}} \| ^2} \big) \big( \mathbf{I}-\frac{\mathbf{h}_{\mathsf{e}}\mathbf{h}_{\mathsf{e}}^{\mathsf{H}}}{\| \mathbf{h}_{\mathsf{e}} \| ^2} \big) \mathbf{h}_{\mathsf{b}}}$. It holds that
			\begin{align}
				\cos \psi=\frac{\left| \| \mathbf{h}_{\mathsf{b}} \| -\frac{| \mathbf{h}_{\mathsf{e}}^{\mathsf{H}}\mathbf{h}_{\mathsf{b}} |^2}{\| \mathbf{h}_{\mathsf{b}} \| \| \mathbf{h}_{\mathsf{e}} \| ^2} \right|^2}{\mathbf{h}_{\mathsf{b}}^{\mathsf{H}}\left( \mathbf{I}-\frac{\mathbf{h}_{\mathsf{e}}\mathbf{h}_{\mathsf{e}}^{\mathsf{H}}}{\left\| \mathbf{h}_{\mathsf{e}} \right\| ^2} \right) \mathbf{h}_{\mathsf{b}}}=\frac{\| \mathbf{h}_{\mathsf{b}} \| ^2\left| 1-\frac{| \mathbf{h}_{\mathsf{e}}^{\mathsf{H}}\mathbf{h}_{\mathsf{b}} |^2}{\| \mathbf{h}_{\mathsf{b}} \| ^2\| \mathbf{h}_{\mathsf{e}} \| ^2} \right|^2}{\| \mathbf{h}_{\mathsf{b}} \| ^2\left( 1-\frac{| \mathbf{h}_{\mathsf{e}}^{\mathsf{H}}\mathbf{h}_{\mathsf{b}} |^2}{\| \mathbf{h}_{\mathsf{b}} \| ^2\| \mathbf{h}_{\mathsf{e}} \| ^2} \right)}.
			\end{align}
			Consequently, we have $\cos\psi=1-\rho$.
		\end{IEEEproof}
		Based on the above discussion, we propose a concept called the \textit{``depth of insecurity''}, which serves as a metric for evaluating the susceptibility of near-field secure transmission in the distance domain. Specifically, when Bob and Eve are located in the same direction, the secrecy capacity decreases as the distance between them shortens. Given $r_\mathsf{b}$, we aim to find an interval $[r_{\min}, r_{\max}]$ of maximum length such that for $r_{\mathsf{e}}\in[r_{\min}, r_{\max}]$, the following condition holds: 
		\begin{align}
			\cos \psi=\frac{| \mathbf{w}_{\mathsf{m}}^{\mathsf{H}}\mathbf{w}_{\infty}^{\star} |^2}{\| \mathbf{w}_{\mathsf{m}} \| ^2\| \mathbf{w}_{\infty}^{\star} \| ^2}=1-\rho\le \Gamma ,
		\end{align}
		where $\Gamma$ is a desired threshold. The depth of insecurity is defined as the length of this interval, i.e., $\mathcal{D}=r_{\max}-r_{\min}$.
		
		When Eve is located in the same direction as Bob but beyond the depth of insecurity, the transmission security is more robustly guaranteed. Consequently, a smaller depth of insecurity signifies better secrecy performance. For clarity, we define the depth of insecurity using a threshold of $\Gamma=\frac{1}{2}$, known as the 3 dB depth. In the sequel, we emply Lemma \ref{lem_angle} to calculate the 3 dB depth under the USW model, which applies to most near-field scenarios. For simplicity, we assume that the UPA is square-shaped, i.e., $M_x=M_z$, and that both users are aligned in the boresight direction, i.e., $(\theta_k,\phi_k)=(\frac{\pi}{2},\frac{\pi}{2})$.
		\vspace{-5pt}
		\begin{lemma}\label{lem_depth}
			The 3 dB depth of insecurity in the boresight direction is given by
			\begin{align}\label{depth}
				\mathcal{D} _{3\mathrm{dB}}=\begin{cases}
					\frac{2r_{\mathsf{b}}^{2}r_{\mathrm{s}}}{r_{\mathsf{s}}^{2}-r_{\mathsf{b}}^{2}}&		r_{\mathsf{b}}<r_{\mathsf{s}}\\
					\infty&		r_{\mathsf{b}}\ge r_{\mathsf{s}}\\
				\end{cases},
			\end{align}
			where $r_{\mathsf{s}}=\frac{Md^2}{4\lambda \Upsilon _{3\mathrm{dB}}^{2}}$, and $\Upsilon _{3\mathrm{dB}}=0.79$.
		\end{lemma}
		\vspace{-2pt}
		\begin{IEEEproof}
			Please refer to Appendix~\ref{proof_lem_depth} for more details.  
		\end{IEEEproof}
		As observed in \eqref{depth}, the depth of insecurity approaches infinity when Bob is located at a distance greater than $r_\mathsf{s}$. Consequently, the region within $r_\mathsf{s}$ is the ``security region'' for Bob, where secure transmission is achievable. Furthermore, we note that increasing the number of transmit antennas can expand the security region and reduce the depth of insecurity, thereby enhancing near-field secure transmission. Specifically, for a given $r_{\mathsf{b}}$, to achieve secure transmission to Bob, the number of antennas must satisfy the condition $M>\frac{4\lambda\Upsilon_{3\mathrm{dB}}r_{\mathsf{b}}}{d^2}\triangleq M_{\mathsf{s}}$.
		\section{Analysis of the Minimum Required Power}\label{mini_Power}
		\subsection{Minimum Power Requirement Characterization}
		For clarity, we denote $t= {\left\| \mathbf{w} \right\| ^2}$ and $\mathbf{v}= t^{-1/2}{\mathbf{w}}$. Given the target secrecy rate $\mathsf{R}_0>0$, the minimum required power and the corresponding beamformer can be found by solving the following problem: 
		\begin{subequations}
			\begin{align}\label{Active_Beamforming_Opt_Power_Min_New}
				\min_{{\mathbf v},t}~~&t\\
				{\rm{s.t.}}~~&\frac{1+t\sigma _{\mathsf{b}}^{-2}\lvert{\mathbf v}^{\mathsf H}{\mathbf h}_{\mathsf{b}}\rvert^2}
				{1+t\sigma _{\mathsf{e}}^{-2}\lvert{\mathbf v}^{\mathsf H}{\mathbf h}_{\mathsf{e}}\rvert^2}\geq2^{{\mathsf{R}_0}},~t> 0,~\lVert{\mathbf v}\rVert^2=1.
			\end{align}
		\end{subequations}
		Multiplying both sides of the first constraint in \eqref{Active_Beamforming_Opt_Power_Min_New} by $1+t\sigma _{\mathsf{e}}^{-2}\lvert{\mathbf v}^{\mathsf H}{\mathbf h}_{\mathsf{e}}\rvert^2$ gives
		\begin{align}\label{P5_Cons1_R2}
			{\mathbf v}^{\mathsf H}((1-2^{{\mathsf{R}_0}}){\mathbf I}+t(\sigma _{\mathsf{b}}^{-2}{\mathbf h}_{\mathsf{b}}{\mathbf h}_{\mathsf{b}}^{\mathsf H}-2^{{\mathsf{R}_0}}\sigma _{\mathsf{e}}^{-2}{\mathbf h}_{\mathsf{e}}{\mathbf h}_{\mathsf{e}}^{\mathsf H})){\mathbf v}\geq 0.
		\end{align}
		By defining ${\bm\Theta}\triangleq\frac{1}{\sigma _{\mathsf{b}}^{2}}{\mathbf h}_{\mathsf{b}}{\mathbf h}_{\mathsf{b}}^{\mathsf H}-\frac{2^{{\mathsf{R}_0}}}{\sigma _{\mathsf{e}}^{2}}{\mathbf h}_{\mathsf{e}}{\mathbf h}_{\mathsf{e}}^{\mathsf H}\in{\mathbbmss C}^{M\times M}$, we have
		\begin{equation}\label{Power_Min_Eigenvalue}
			\begin{split}
				{\mathbf v}^{\mathsf H}((1-2^{{\mathsf{R}_0}}){\mathbf I}+t{\bm\Theta}){\mathbf v} \le 1-2^{{\mathsf{R}_0}}+t\mu_{\bm \Theta },
			\end{split}
		\end{equation}
		where $\mu_{\bm \Theta }$ denotes the principal eigenvalue of ${\bm\Theta}$. Upon combining \eqref{P5_Cons1_R2} and \eqref{Power_Min_Eigenvalue}, we find that $t$ must satisfy $t\geq\frac{2^{{\mathsf{R}_0}}-1}{\mu_{\bm \Theta }}>0$ to guarantee the feasibility of problem \eqref{Active_Beamforming_Opt_Power_Min_New}. This means that the minimum value of $t$, i.e., the minimum required power, can be written as follows:
		\begin{align}
			\mathsf{P}= \frac{2^{{\mathsf{R}_0}}-1}{\mu_{\bm \Theta }}.  
		\end{align}
		In this case, the corresponding $\mathbf v$ is the normalized principal eigenvector of ${\bm\Theta}$. By applying the matrix determinant lemma, we deduce the following theorem.
		\vspace{-5pt}
		\begin{theorem}\label{theorem2}
			The minimum required transmit power to guarantee a target secrecy rate $\mathsf{R}_0$ is given by
			\begin{equation}
				\mathsf{P}=\frac{2\left( 2^{\mathsf{R}_0}-1 \right)}{\xi +\sqrt{\xi ^2+\chi}},\label{Optimal_Active_Beamforming_Transmit_Power}
			\end{equation}
			where $\xi =\sigma _{\mathsf{b}}^{-2}G_{\mathsf{b}}-2^{\mathsf{R}_0}\sigma _{\mathsf{e}}^{-2}G_{\mathsf{e}}$ and $\chi =2^{\mathsf{R}_0+2}\sigma _{\mathsf{b}}^{-2}\sigma _{\mathsf{e}}^{-2}G_{\mathsf{e}}G_{\mathsf{b}}\left( 1-\rho \right) $.
		\end{theorem}
		\vspace{-5pt}
		\begin{IEEEproof}
			Please refer to Appendix \ref{Appendix3} for more details.	
		\end{IEEEproof}
		\vspace{-5pt}
		\begin{corollary}\label{cor_unachievable}
			The finite target secrecy rate $\mathsf{R}_0$ is unachievable if and only if $\rho=1$ and $\frac{\sigma _{\mathsf{b}}^{-2}G_{\mathsf{b}}}{\sigma _{\mathsf{e}}^{-2}G_{\mathsf{e}}}\le2^{\mathsf{R}_0}$.
		\end{corollary}
		\vspace{-5pt}
		\begin{IEEEproof}
			The unachievability of $\mathsf{R}_0$ means that $\mathsf{P}=\infty$, which yields $\xi +\sqrt{\xi ^2+\chi}=0$. In this case, we have $\chi=0$ and $\xi\le0$, and it follows that $\rho=1$ and $\frac{\sigma _{\mathsf{b}}^{-2}G_{\mathsf{b}}}{\sigma _{\mathsf{e}}^{-2}G_{\mathsf{e}}}\le2^{\mathsf{R}_0}$. The final results follow immediately.
		\end{IEEEproof}
		Based on the previous results and the above corollary, we have the following observation.
			\vspace{-5pt}
			\begin{remark}\label{rem_p_compare_1}
				Unlike FFC, NFC can achieve any desired secrecy rate for Bob, regardless of his location, provided there is sufficient transmit power. 	
			\end{remark}
			
			\begin{table}[!t]
				\center
				\scalebox{0.77}{
					\begin{tabular}{|c|c|}
							\hline
							\textbf{Model} & \textbf{Minimum Required Power} \\ \hline
							\textbf{UPW}   & $\frac{2\left( 2^{\mathsf{R}_0}-1 \right)}{\frac{MA\Psi _{\mathsf{b}}}{4\pi r_{\mathsf{b}}^{2}\sigma _{\mathsf{b}}^{2}}-\frac{2^{\mathsf{R}_0}MA\Psi _{\mathsf{e}}}{4\pi r_{\mathsf{e}}^{2}\sigma _{\mathsf{e}}^{2}}+\sqrt{\left( \frac{MA\Psi _{\mathsf{b}}}{4\pi r_{\mathsf{b}}^{2}\sigma _{\mathsf{b}}^{2}}-\frac{2^{\mathsf{R}_0}MA\Psi _{\mathsf{e}}}{4\pi r_{\mathsf{e}}^{2}\sigma _{\mathsf{e}}^{2}} \right) ^2+\frac{2^{\mathsf{R}_0+2}M^2A^2\Psi _{\mathsf{b}}\Psi _{\mathsf{e}}}{\left( 4\pi \right) ^2r_{\mathsf{b}}^{2}r_{\mathsf{e}}^{2}\sigma _{\mathsf{b}}^{2}\sigma _{\mathsf{e}}^{2}}\left( 1-\rho _{_{\mathsf{P}}} \right)}}$                \\ \hline
							\textbf{USW}   & $\frac{2\left( 2^{\mathsf{R}_0}-1 \right)}{\frac{MA\Psi _{\mathsf{b}}}{4\pi r_{\mathsf{b}}^{2}\sigma _{\mathsf{b}}^{2}}-\frac{2^{\mathsf{R}_0}MA\Psi _{\mathsf{e}}}{4\pi r_{\mathsf{e}}^{2}\sigma _{\mathsf{e}}^{2}}+\sqrt{\left( \frac{MA\Psi _{\mathsf{b}}}{4\pi r_{\mathsf{b}}^{2}\sigma _{\mathsf{b}}^{2}}-\frac{2^{\mathsf{R}_0}MA\Psi _{\mathsf{e}}}{4\pi r_{\mathsf{e}}^{2}\sigma _{\mathsf{e}}^{2}} \right) ^2+\frac{2^{\mathsf{R}_0+2}M^2A^2\Psi _{\mathsf{b}}\Psi _{\mathsf{e}}}{\left( 4\pi \right) ^2r_{\mathsf{b}}^{2}r_{\mathsf{e}}^{2}\sigma _{\mathsf{b}}^{2}\sigma _{\mathsf{e}}^{2}}\left( 1-\rho _{_{\mathsf{S}}} \right)}}$                \\ \hline
							\textbf{NUSW}  & $\frac{2\left( 2^{\mathsf{R}_0}-1 \right)}{\sigma _{\mathsf{b}}^{-2}G_{\mathsf{b}}^{\mathsf{N}}-2^{\mathsf{R}_0}\sigma _{\mathsf{e}}^{-2}G_{\mathsf{e}}^{\mathsf{N}}+\sqrt{\left( \sigma _{\mathsf{b}}^{-2}G_{\mathsf{b}}^{\mathsf{N}}-2^{\mathsf{R}_0}\sigma _{\mathsf{e}}^{-2}G_{\mathsf{e}}^{\mathsf{N}} \right) ^2+2^{\mathsf{R}_0+2}\sigma _{\mathsf{b}}^{-2}\sigma _{\mathsf{e}}^{-2}G_{\mathsf{b}}^{\mathsf{N}}G_{\mathsf{e}}^{\mathsf{N}}\left( 1-\rho _{_{\mathsf{N}}} \right)}}$                \\ \hline
				\end{tabular}}
				\caption{Minimum required transmission power under different channel models.}
				\vspace{-7pt}
				\label{table_p}
			\end{table}
			
			Note that minimum required power $\mathsf{P}$ is also expressed as a function of the channel gains and the channel correlation factor, which were derived in the previous section. We conclude the expressions of the minimum required power for different channel models in Table~\ref{table_p}.  
			
			In the following subsection, we focus on the asymptotic analysis as $M_x,M_z\rightarrow\infty$ ($M\rightarrow\infty$) for the various far-field and near-field channel models to unveil further insights. For brevity, we assume $\sigma_\mathsf{b}^2=\sigma_\mathsf{e}^2=\sigma^2$.

		\subsection{Far-Field and Near-Field Power Scaling}
		\subsubsection{UPW Model}
		We commence with FFC by studying the UPW model. 
		\vspace{-5pt}
		\begin{corollary}\label{cor_upw_minP}
			When $M\rightarrow\infty$, the minimum required power under the UPW model satisfies
			\begin{align}
				\lim_{M\rightarrow\infty}\mathsf{P}= \begin{cases}
					\infty&		\left( \theta _{\mathsf{b}},\phi _{\mathsf{b}} \right) =\left( \theta _{\mathsf{e}},\phi _{\mathsf{e}} \right) \& r_\mathsf{e}\le2^{{\mathsf{R}_0}/2}r_\mathsf{b}\\
					0&		\text{else}\\
				\end{cases}.
			\end{align}  
		\end{corollary}
		\begin{IEEEproof}
			Please refer to Appendix \ref{proof_cor_upw_minP} for more details.
		\end{IEEEproof}
		From Corollary \ref{cor_upw_minP}, we can draw the following remark.
		\vspace{-5pt}
		\begin{remark}\label{rem_P_upw}
			The minimum required power under the UPW model may drop to zero as the number of transmit antennas increases indefinitely. This means that the target secrecy rate can be achieved with no transmit power, which is impractical.
		\end{remark}
		
		\subsubsection{USW Model}
		We next shift to the NFC.
		\vspace{-5pt}
		\begin{corollary}\label{cor_usw_minP}
			When $M\rightarrow\infty$, the minimum required power under the USW model is given by zero, i.e., $\lim_{M\rightarrow\infty}\mathsf{P}=0$.  
		\end{corollary}
		\vspace{-5pt}
		\begin{IEEEproof}
			For the USW model, we have $\lim_{M\rightarrow\infty} \rho_{_{\mathsf{S}}} = 0$. Then the results can be obtained by following the steps to obtain Corollary \ref{cor_upw_minP}.
		\end{IEEEproof}
		\begin{remark}\label{rem_P_usw}
			The results of Corollary \ref{cor_usw_minP} suggest that under the USW model, when the number of transmit antennas approaches infinity, the target secrecy rate can be achieved with zero transmit power, which contradicts the principle of energy conservation.
		\end{remark}

		\subsubsection{NUSW Model}
		Turn now to the NFC case under the NUSW model.
		\vspace{-5pt}
		\begin{corollary}\label{cor_nusw_minP}
			When $M_x,M_z\rightarrow\infty$, the minimum required power under the NUSW model satisfies
			\begin{align}
				\lim_{M_x,M_z\rightarrow\infty}\mathsf{P}\approx\frac{2( 2^{\mathsf{R}_0}-1) d^2}{\sigma^{-2}A}.
			\end{align}
		\end{corollary}
		\vspace{-2pt}
		\begin{IEEEproof}
			Under the NUSW model, we have $\lim_{M_x,M_z\rightarrow\infty} \rho_{_{\mathsf{N}}} \ll 1$ and $\lim_{M_x,M_z\rightarrow\infty} G_{\mathsf{b}}^{\mathsf{N}}=\frac{A}{2d^2}$. Following the steps outlined in Appendix \ref{proof_cor_upw_minP}, we have $\lim_{M_x,M_z\rightarrow \infty} \mathsf{P}=\lim_{M_x,M_z\rightarrow \infty} \frac{2^{\mathsf{R}_0}-1}{\sigma ^{-2}G_{\mathsf{b}}}=\frac{2\left( 2^{\mathsf{R}_0}-1 \right) d^2}{\sigma ^{-2}A}$.
		\end{IEEEproof}

		\begin{remark}\label{rem_P_nusw}
			Corollary \ref{cor_usw_minP} suggests that, compared to the UPW and USW models, the minimum required power under the NUSW model decreases to a lower bound greater than zero as the number of transmit antennas increases.
		\end{remark}
		By comparing the results of the different channel models, we can find that the NUSW model is more realistic than the other models when investigating power scaling, as it adheres to the principle of energy conservation.
		
			\begin{figure}[!t]
			\centering
			\includegraphics[height=0.28\textwidth]{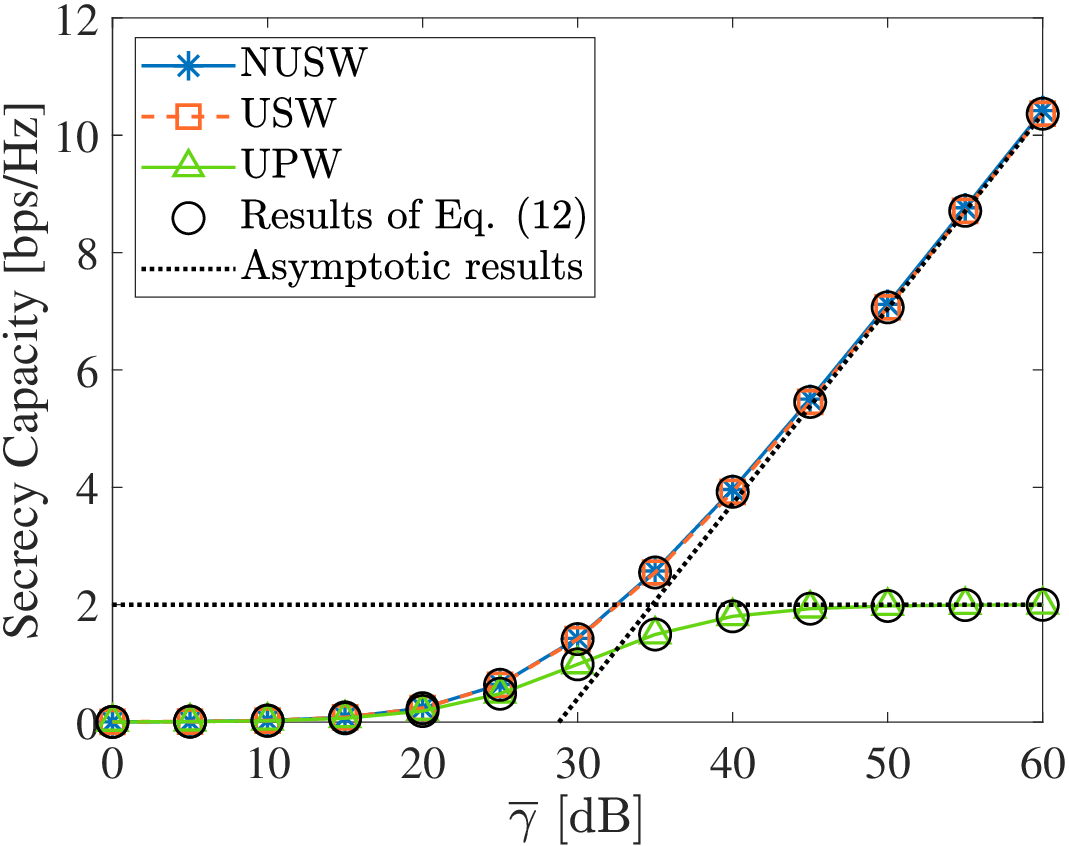}
			\caption{Secrecy capacity versus $\overline{\gamma}$.}
			\label{C_p}
			\vspace{-5pt}
		\end{figure}
		
		\begin{figure}[!t]
			\centering
			\includegraphics[height=0.28\textwidth]{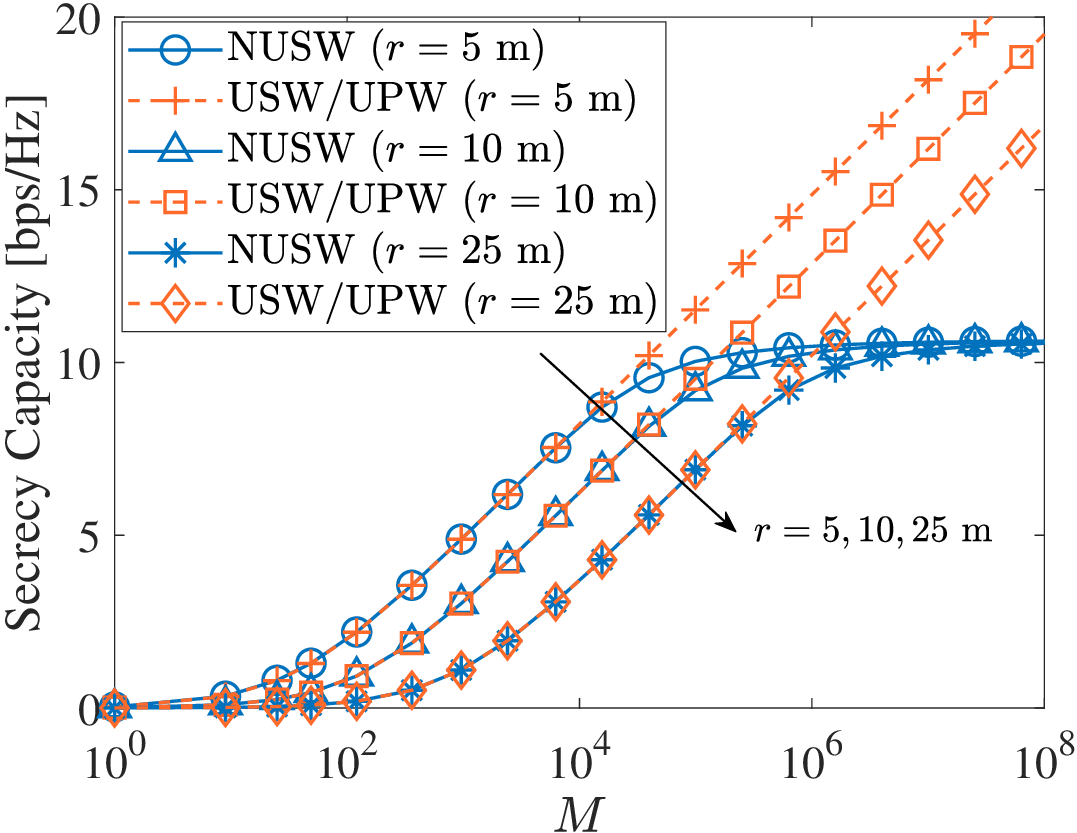}
			\caption{Secrecy capacity versus $M$, with $(\theta_\mathsf{e},\phi_\mathsf{e})=(\frac{2\pi}{3},\frac{\pi}{3})$.}
			\label{C_M}
			\vspace{-5pt}
		\end{figure}

		\section{Numerical Results}\label{numerical}
		In this section, numerical results are presented to demonstrate the secrecy performance in both the near field and far field, and to verify the accuracy of the analytical results. Unless otherwise specified, the simulation parameters are set as follows: $\lambda=0.125$ m, $d=\frac{\lambda}{2}$, $A=\frac{\lambda ^2}{4\pi}$, $M_x=M_z=51$, $\overline{\gamma}=40$ dB, $\sigma^2=-10$ dB, $\mathsf{R}_0=1$ bps/Hz, $(\theta_\mathsf{b},\phi_\mathsf{b})=(\theta_\mathsf{e},\phi_\mathsf{e})=(\frac{\pi}{3},\frac{2\pi}{3})$, $r_\mathsf{b}=r$ and $r_\mathsf{e}=2r$ with $r=10$ m, and $T=100$.
		\subsection{Secrecy Capacity}
		{\figurename} {\ref{C_p}} plots the secrecy capacity in terms of the transmit SNR. The derived closed-form results match well with the simulation results as per the definition given as \eqref{Secrecy_Channel_Capacity}, and the asymptotic results precisely track the trends observed in the high-SNR region. Additionally, we can observe that when Eve is located in the same direction with Bob, the secrecy capacity of the UPW model converges to an upper bound of $2\log_2\frac{r_\mathsf{e}}{r_\mathsf{b}}=2$, while those of the near-field channel models keep increasing with the SNR. In other words, the high-SNR slope of the secrecy capacity achieved by NFC is larger than that achieved by FFC, which verifies our analysis.

		{\figurename} {\ref{C_M}} plots the secrecy capacity as a function of the number of transmit antennas for various values of $r$. It can be observed that for small values of $M$, the secrecy capacities under all channel models increase linearly with $\log M$. This occurs because, when the antenna number is small, the users can be treated as in the far-field region, where all models are sufficiently accurate. However, as $M$ grows larger, the variations in channel powers and phases across the array become pronounced. In such cases, the secrecy capacities under the UPW and USW models are overestimated due to neglect of these variations in wave propagation, and they will increase unboundedly with $M$, breaking the energy-conservation laws. This is in line with the statements in Remarks~\ref{rem_upw_M} and \ref{rem_usw_M}. In contrast, as $M$ increases, the secrecy capacity under the NUSW model is capped at finite values, aligning with the results in Remark \ref{rem_nusw_M} and justifying the superior accuracy and robustness of the NUSW model.
		
		\begin{figure}[!t]
			\centering
			\includegraphics[height=0.28\textwidth]{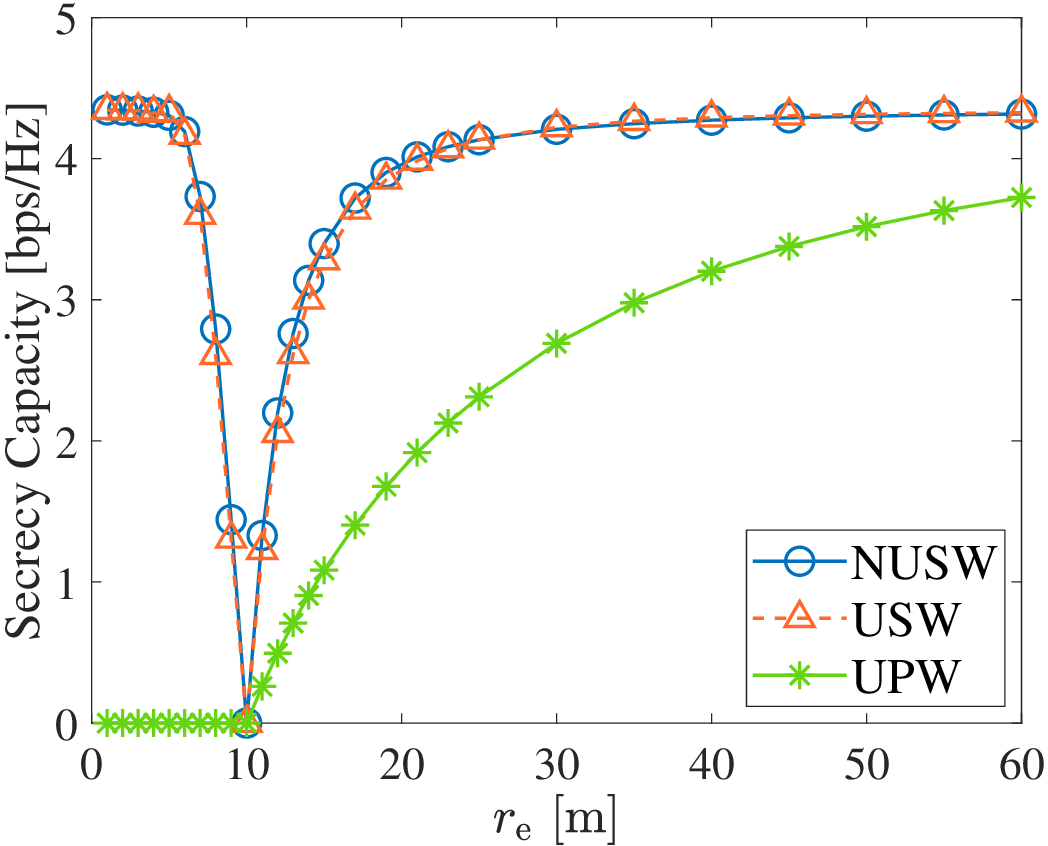}
			\caption{Secrecy capacity versus $r_\mathsf{e}$ when $r_\mathsf{b}=10$ m.}
			\label{C_re}
			\vspace{-7pt}
		\end{figure}

		\begin{figure*}[!t]
			\centering
			\subfigbottomskip=5pt
			\subfigcapskip=0pt
			\setlength{\abovecaptionskip}{0pt}
			\subfigure[NUSW.]
			{
				\includegraphics[height=0.25\textwidth]{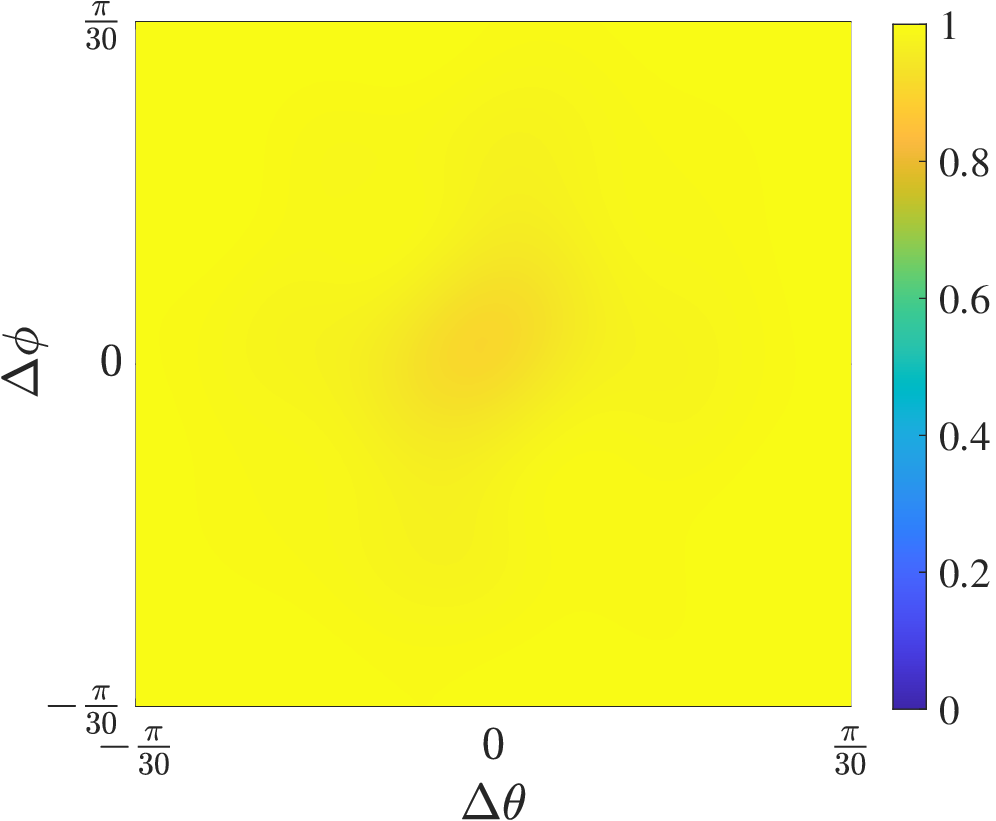}
				\label{per_nusw}	
			}
			\quad
			\subfigure[USW.]
			{
				\includegraphics[height=0.25\textwidth]{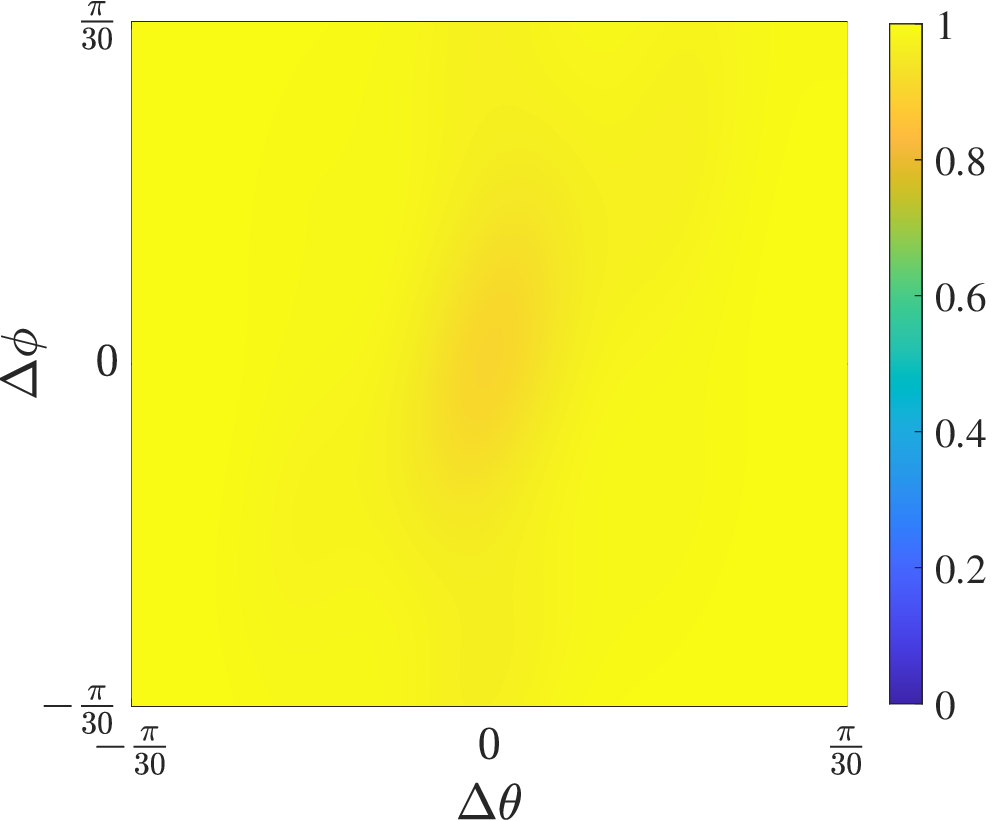}
				\label{per_usw}	
			}
			\quad
			\subfigure[UPW.]
			{
				\includegraphics[height=0.25\textwidth]{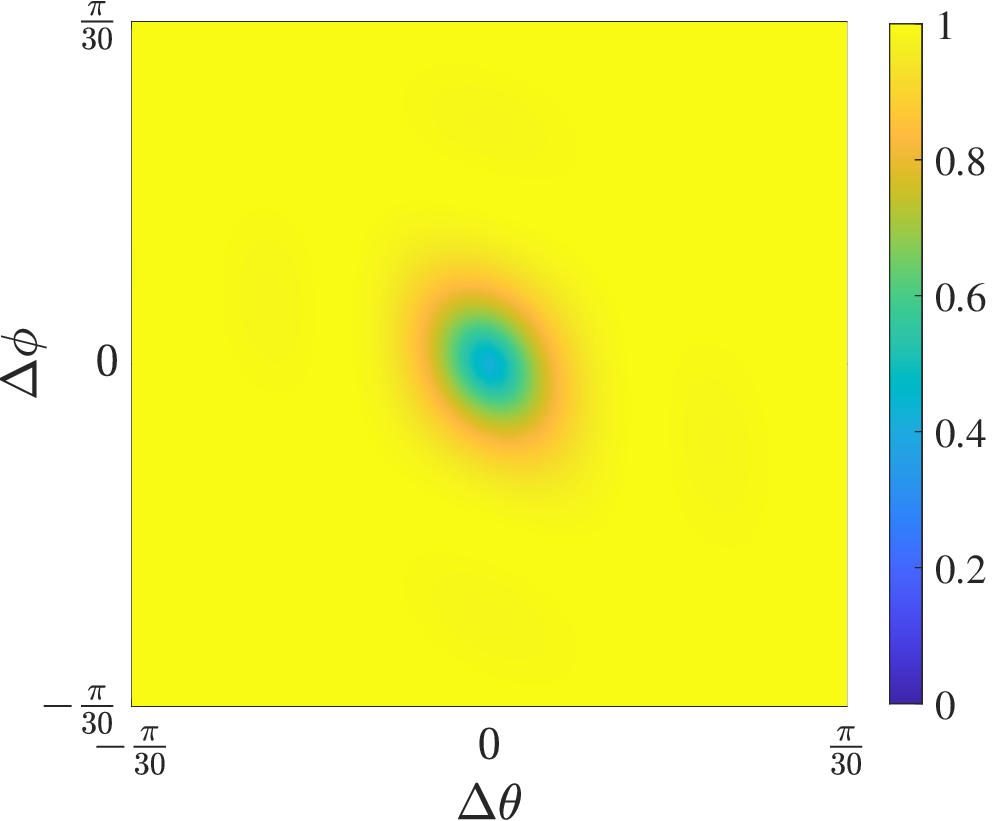}
				\label{per_upw}	
			}
			\caption{Normalized secrecy capacity versus small angular difference between Bob and Eve.}
			\vspace{-5pt}
			\label{per}
		\end{figure*}
		
		{\figurename} {\ref{C_re}} illustrates the secrecy channel capacity with respect to the distance of Eve for various channel models, where Bob and Eve are located in the same direction with $(\theta_\mathsf{b},\phi_\mathsf{b})=(\theta_\mathsf{e},\phi_\mathsf{e})=(\frac{\pi}{3},\frac{2\pi}{3})$. As can be seen, when Eve is closer to the BS than Bob, its channel condition poorer will be worse than Bob's, and the far-field secrecy capacity remains zero, making secure transmission within the far field unachievable. By contrast, secure transmission under NFC is always feasible except when Bob and Eve are at the exact same location. These behaviors are consistent with the insights provided in Remarks \ref{rem_upw_C} and \ref{rem_usw_C}, underscoring the superior secrecy performance of NFC. Moreover, we can observe that NFC consistently achieves higher secrecy capacity than FFC, even when Bob is farther from the BS than Eve. This advantage is attributed to the enhanced distance resolution offered by the spherical-wave model, which enables more precise focusing of signal energy on Bob while minimizing leakage to Eve. However, this gap decreases as $r_{\mathsf{e}}$ increase. This is because the near-field effect is more pronounced at short distances and diminishes as Bob or Eve moves toward the far-field region.
		
		While we find that the secrecy capacity for co-directional Bob and Eve improves under NFC compared to the FF case, it is nearly impossible for them to be positioned in exactly the same direction in real-word scenarios, typically only similarly. Therefore, it would be important to explore how small angular perturbations between Bob and Eve impact the secrecy capacity. By letting $(\theta_{\mathsf{e}},\phi_{\mathsf{e}})=(\theta_{\mathsf{b}}+\Delta\theta,\phi_{\mathsf{b}}+\Delta\phi)$ with $(\Delta\theta,\Delta\phi)$ denoting the small angular perturbations, Fig.~\ref{per} illustrates how secrecy capacity fluctuates with changes in $(\Delta\theta,\Delta\phi)$. The results show that the capacities for the NUSW and USW models remain high regardless of the angle differences. On the other hand, while the secrecy capacity under the UPW model is comparable to NFC when Bob and Eve are in different directions, it significantly deteriorates when their angular difference is minimal. These observations confirm the superior performance of NFC in scenarios where Eve is aligned with Bob or even with small angular deviations.
		
		\begin{figure}[!t]
			\centering
			\includegraphics[height=0.28\textwidth]{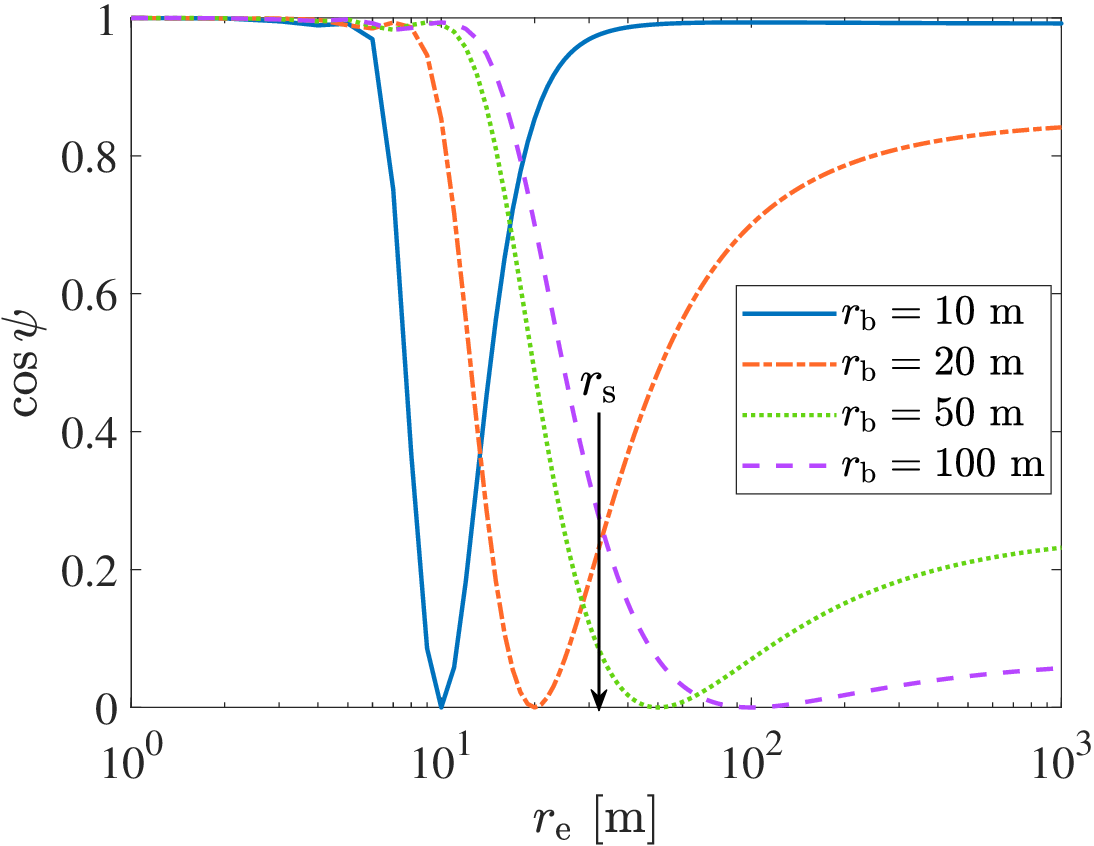}
			\caption{$\cos \psi$ versus $r_\mathsf{e}$.}
			\label{depth_re}
			\vspace{-7pt}
		\end{figure}
		
		\begin{figure}[!t]
			\centering
			\includegraphics[height=0.28\textwidth]{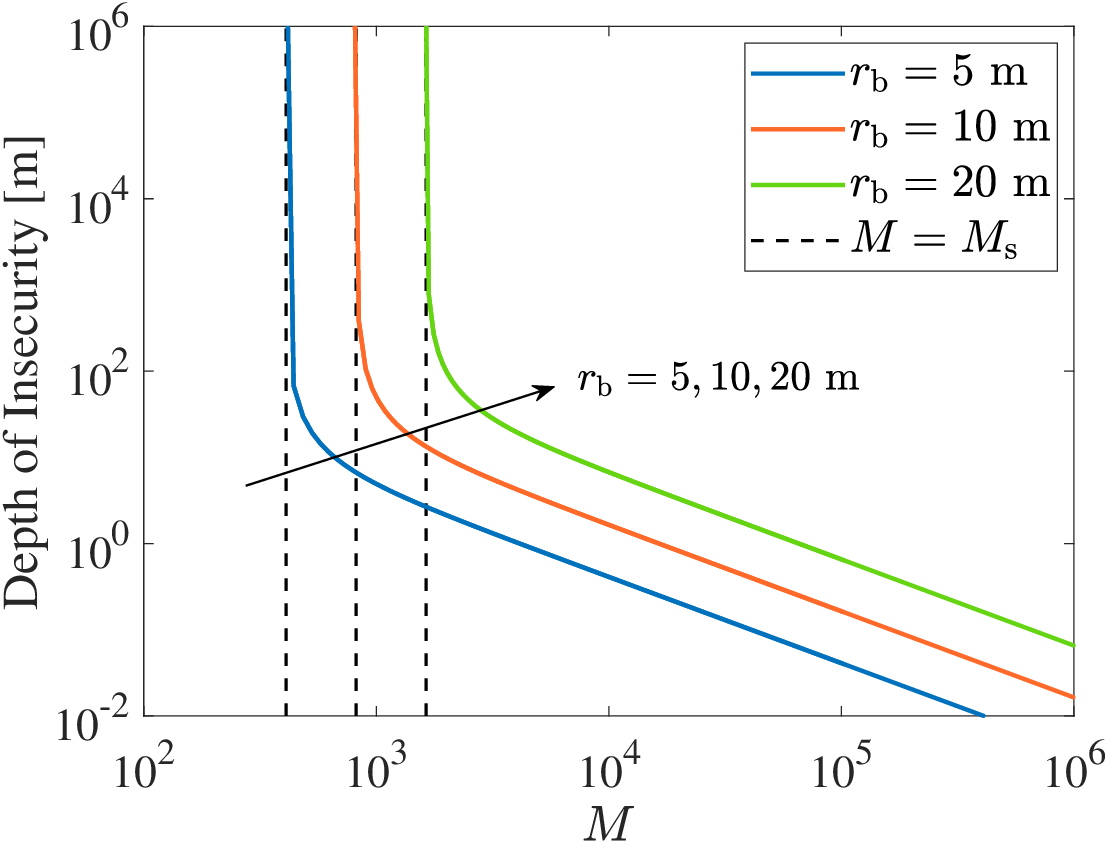}
			\caption{3 dB depth of insecurity versus $M$.}
			\label{depth_M}
			\vspace{-5pt}
		\end{figure}
		
		{\figurename} {\ref{depth_re}} plots $\cos\psi$ as a function of $r_\mathsf{e}$ for different $r_\mathsf{b}$ values to illustrate the depth of insecurity. It can be observed that the 3 dB depth (obtained by studying $\cos\psi\le\frac{1}{2}$) becomes infinite when $r_\mathsf{b}\ge r_\mathsf{s}$, which is consistent with the results of \eqref{depth}. {\figurename} {\ref{depth_M}} illustrates the 3 dB depth versus the number of transmit antennas. In line with our prior analyses, the numerical results show that the depth of insecurity decreases as $M$ increases. The dashed lines represent the case when the antenna numbers is $M_\mathsf{s}$, leading to $r_\mathsf{s}= r_\mathsf{b}$. Therefore, when $M\le M_\mathsf{s}$, the depth is infinite. Additionally, we note that for a given $M$, the depth of insecurity is positively correlated with Bob's distance from the BS, which means that secure transmission is more readily achieved when the intended receiver is closer to the transmitter.
		
		\begin{figure}[!t]
			\centering
			\includegraphics[height=0.28\textwidth]{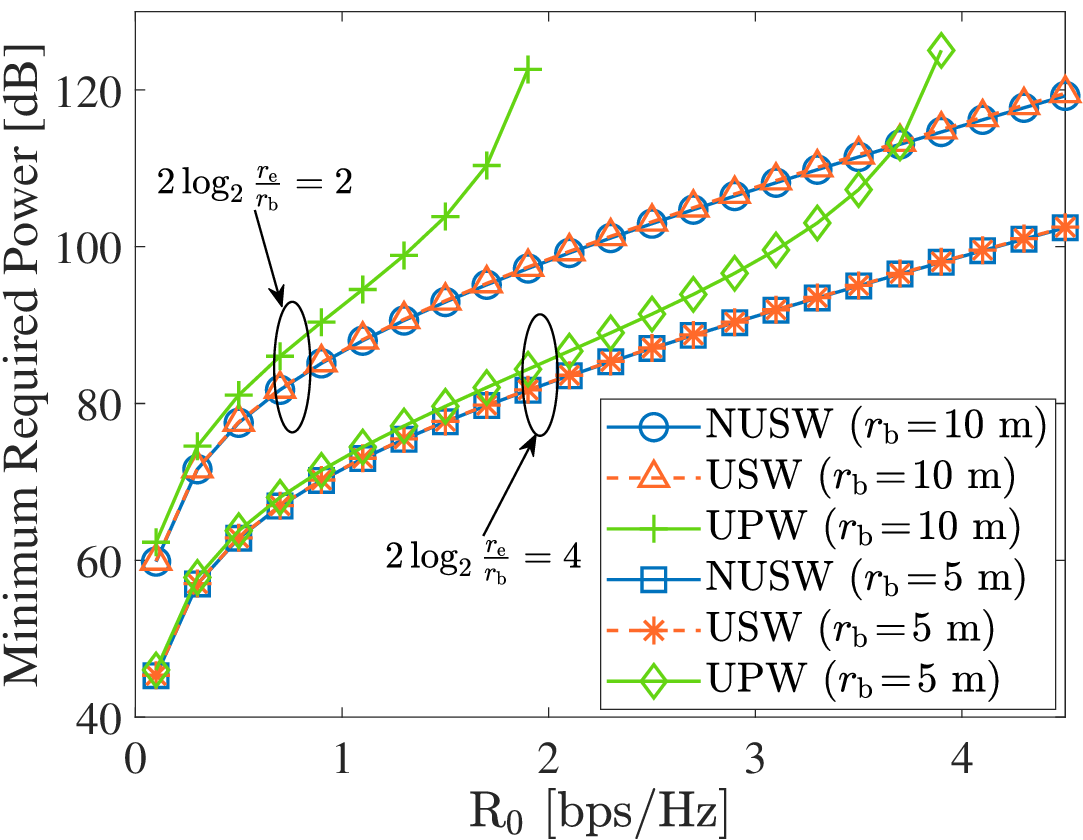}
			\caption{Minimum required power versus $\mathsf{R}_0$.}
			\label{P_r0}
			\vspace{-5pt}
		\end{figure}

		\subsection{Minimum Required Power}
\begin{figure}[!t]
			\centering
			\includegraphics[height=0.28\textwidth]{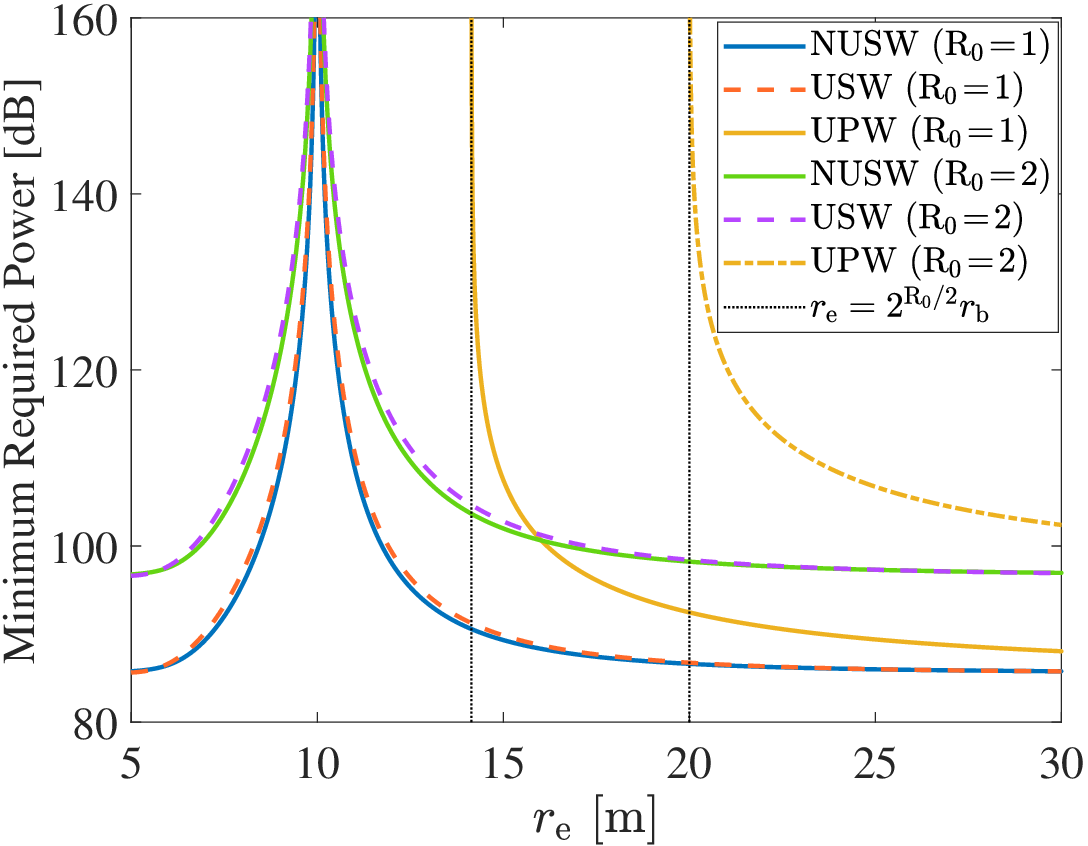}
			\caption{Minimum required power versus $r_\mathsf{e}$ with $r_\mathsf{b}=10$ m.}
			\label{P_re}
			\vspace{-5pt}
		\end{figure}

		Fig.~\ref{P_r0} illustrates the minimum required power for various channel models to achieve different target secrecy rates when Bob and Eve are located in the same direction. It can be observed that the UPW model requires more transmission power than the USW and NUSW models to achieve a same target secrecy rate, with the disparity increasing as $\mathsf{R}_0$ grows. Notably, when the target secrecy rate exceeds the threshold $\mathsf{R}_0>2\log_2\frac{r_\mathsf{e}}{r_\mathsf{b}}$, the UPW model becomes incapable of achieving it, regardless of the available transmission power. Conversely, the near-field model can attain arbitrarily high secrecy rates, given sufficient transmission power. These findings are consistent with the discussions in Remark~\ref{rem_p_compare_1}. Additionally, {\figurename} {\ref{P_re}} shows the minimum required power as a function of Eve's distance from the BS. It is worth noting that the under FFC, a target secrecy rate is only achievable when $r_\mathsf{e}>2^{\mathsf{R}_0/2}r_\mathsf{b}$, whereas NFC can achieve the target rate as long as $r_\mathsf{e}\ne r_\mathsf{b}$. The above observations demonstrate the advantages of NFC in enhancing physical layer security.
		
		\begin{figure}[!t]
			\centering
			\includegraphics[height=0.28\textwidth]{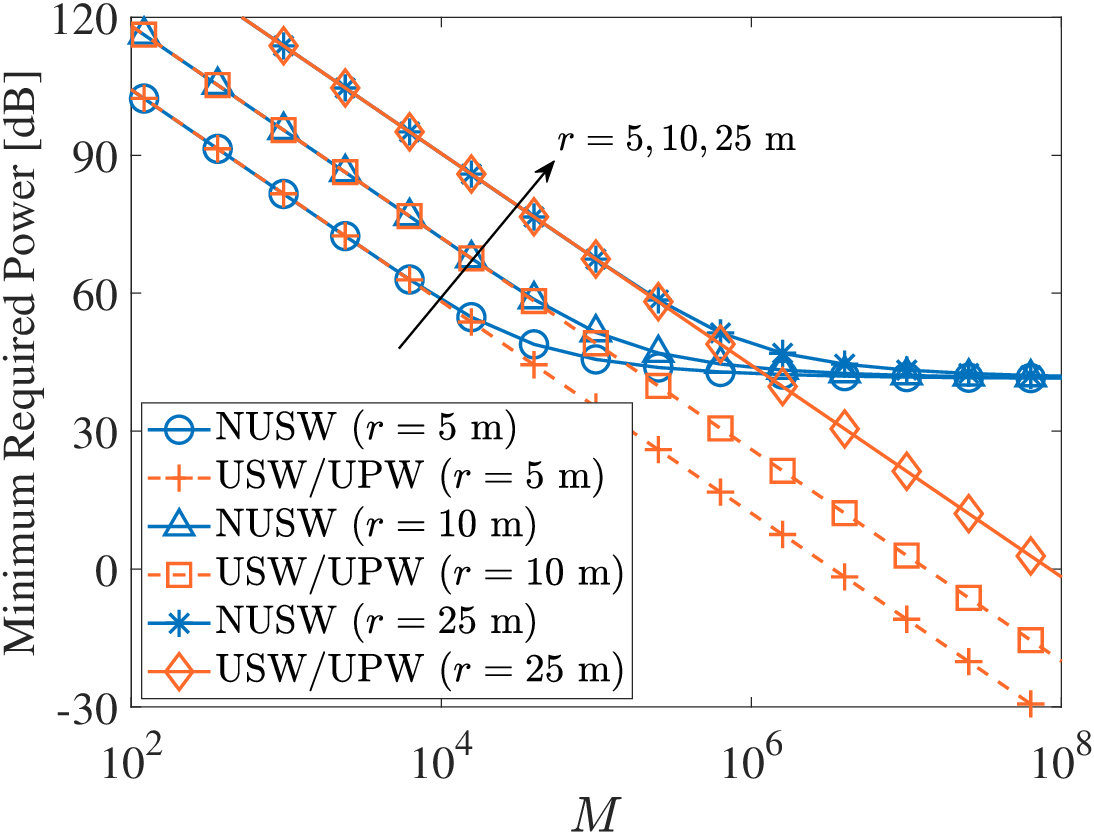}
			\caption{Minimum required power versus $M$, with $(\theta_\mathsf{e},\phi_\mathsf{e})=(\frac{2\pi}{3},\frac{\pi}{3})$.}
			\label{P_M}
			\vspace{-5pt}
		\end{figure}
		
		In {\figurename} {\ref{P_M}}, we present the minimum required power as a function of the antenna number $M$. Initially, for smaller values of $M$, the minimum required power under all models approximately decreases linearly with $\log M$. However, as $M$ increases, expanding the near-field region, the UPW and USW models become less accurate. Specifically, the minimum required power under the UPW/USW models continues to decrease and approaches zero, which is unrealistic in practical scenarios. In contrast, for the more accurate NUSW model, the minimum required power converges to a lower bound greater than zero, reflecting a more realistic scenario. These observations corroborate the insights provided in Remarks \ref{rem_P_upw}, \ref{rem_P_usw} and \ref{rem_P_nusw}. 
		
		\section{Conclusion}\label{conclusion}
		This article has analyzed the performance of physical layer security in terms of secrecy capacity and minimum required power. Using various commonly employed far-field and near-field channel models, namely the UPW, USW, and NUSW models, we derived closed-form expressions for the secrecy capacity under a power budget and the minimum power required to achieve a target secrecy rate. To gain further insights, we conducted asymptotic analysis to unveil the capacity scaling law and power scaling law, and compared the secrecy performance across different models. Additionally, we introduced the concept of the depth of insecurity as a metric to evaluate the effectiveness of near-field beamfocusing in enhancing secrecy performance. Through both theoretical analyses and numerical simulations, we demonstrated that NFC offers superior secrecy performance, owing to spherical-wave EM propagation, compared to FFC, thereby enhancing physical layer security.
		
		\begin{appendix}
			\setcounter{equation}{0}
			\renewcommand\theequation{A\arabic{equation}}
			\subsection{Proof of Theorem \ref{Theorem_Secrecy_Capacity}}\label{Proof_Theorem_Secrecy_Capacity}
			The eigenvalues of ${\bm\Delta}$ can be calculated from the characteristic equation $\det({\bm\Delta}-\mu{\mathbf I})=0$. Employing the fact of ${\mathbf{Q}}_{\mathsf{e}}\succ{\mathbf{I}}$, we obtain $\det(\mathbf{\Delta }-\mu \mathbf{I})=0\Leftrightarrow\det({\mathbf{Q}}_{\mathsf{e}}^{{1}/{2}})\det(\mathbf{\Delta }-\mu \mathbf{I})\det({\mathbf{Q}}_{\mathsf{e}}^{{1}/{2}})=0$, i.e., $\det(\mu \mathbf{Q}_{\mathsf{e}}^{1/2}\mathbf{IQ}_{\mathsf{e}}^{1/2}-\mathbf{Q}_{\mathsf{e}}^{1/2}\mathbf{\Delta Q}_{\mathsf{e}}^{1/2})$. Upon defining $\mathbf{\Sigma }\triangleq \left( \mu -1 \right) \mathbf{I}+\mu \overline{\gamma }_{\mathsf{e}}\mathbf{h}_{\mathrm{e}}\mathbf{h}_{\mathrm{e}}^{\mathsf{H}}$, we have
			\begin{align}\label{Ch_Poly_1}
				\det ( {\bm\Sigma}-\overline{\gamma}_\mathsf{b}\mathbf{{h}}_{\mathsf{b}}\mathbf{{h}}_{\mathsf{b}}^{\mathsf H} )=0.
			\end{align}
			Leveraging the matrix determinant lemma \cite{horn2012matrix}, we can rewrite \eqref{Ch_Poly_1} as follows: 
			\begin{align}\label{Ch_Poly_2}
				(1-\overline{\gamma}_\mathsf{b}\mathbf{{h}}_{\mathsf{b}}^{\mathsf H}{\bm\Sigma}^{-1}\mathbf{{h}}_{\mathsf{b}})\det({\bm\Sigma})=0,
			\end{align}
			where ${\bm\Sigma}^{-1}$ can be calculated using the Woodbury matrix identity \cite{horn2012matrix}, which yields
			\begin{align}\label{Ch_Poly_3}
				\mathbf{\Sigma}^{-1}=\frac{1}{\mu -1}\mathbf{I}-\frac{\mu \left( \mu -1 \right) ^{-1}\overline{\gamma }_{\mathsf{e}}\mathbf{h}_{\mathrm{e}}\mathbf{h}_{\mathrm{e}}^{\mathsf{H}}}{\mu -1+\mu \overline{\gamma }_{\mathsf{e}}\left\| \mathbf{h}_{\mathrm{e}} \right\| ^2}.
			\end{align}
			It follows from the Sylvester's determinant identity \cite{horn2012matrix} that
			\begin{align}\label{Ch_Poly_4}
				\det(\mathbf{\Sigma})=\left( \mu -1 \right) ^{M-1}\big( \mu -1+\mu \overline{\gamma }_{\mathsf{e}}\left\| \mathbf{h}_{\mathrm{e}} \right\| ^2 \big) .
			\end{align}
			Substituting \eqref{Ch_Poly_3} and \eqref{Ch_Poly_4} into \eqref{Ch_Poly_2} gives
			\begin{equation}\label{Ch_Poly_5}
				\begin{split}
					&\!\!\!( \mu -1 ) ^{M-2}[ \left( 1+\overline{\gamma }_{\mathsf{e}}G_{\mathsf{e}} \right) \left( \mu -1 \right) ^2\!-( \overline{\gamma }_{\mathsf{b}}G_{\mathsf{b}}-\overline{\gamma }_{\mathsf{e}}G_{\mathsf{e}}+\\
					&\!\!\!\overline{\gamma }_{\mathsf{e}}\overline{\gamma }_{\mathsf{b}}G_{\mathsf{b}}G_{\mathsf{e}}\left( 1-\rho \right) ) \left( \mu -1 \right) -\overline{\gamma }_{\mathsf{e}}\overline{\gamma }_{\mathsf{b}}G_{\mathsf{b}}G_{\mathsf{e}}\left( 1-\rho \right) ] =0.
				\end{split}
			\end{equation}
			Let $\{\mu_m\}_{m=1}^{M}$ denote the $M$ roots of \eqref{Ch_Poly_5}. By the quadratic-root formula, we have $\mu_2=\ldots=\mu_{M-1}=0$ and
			\begin{align}
				\mu_1=1+\frac{\alpha\!+\!\sqrt{\alpha^2+4\beta}}{2(1+{\overline{\gamma}_{\mathsf{e}}}G_{\mathsf{e}})},~\mu_M=1+\frac{\alpha\!-\!\sqrt{\alpha^2+4\beta}}{2(1+{\overline{\gamma}_{\mathsf{e}}}G_{\mathsf{e}})}.
			\end{align}
			We note that $(\mu_1-1)(\mu_M-1)=\frac{-\beta}{(1+{\overline{\gamma}_{\mathsf{e}}}\lVert \mathbf{{h}}_{\mathsf{e}} \rVert ^2)^2}\leq0$, which implies $\mu_1\geq\mu_2=\ldots=\mu_{M-1}\geq\mu_M$. Therefore, the principal eigenvalue of $\bm\Delta$ is $\mu_1$, i.e., $\mu_{\bm\Delta}=\mu_1$. Then the final results follow immediately.
			\subsection{Proof of Corollary \ref{cor_upw_p}}\label{proof_cor_p}
			When $\left( \theta _{\mathsf{b}},\phi _{\mathsf{b}} \right) =\left( \theta _{\mathsf{e}},\phi _{\mathsf{e}} \right)$, the results are obtained from \eqref{upw_same} using $\lim_{\overline{\gamma}\rightarrow\infty}\frac{1+{\overline{\gamma}}G_k^{\mathsf{P}}}{{\overline{\gamma}}G_k^{\mathsf{P}}}=1$. 
			
			When $\left( \theta _{\mathsf{b}},\phi _{\mathsf{b}} \right) \ne \left( \theta _{\mathsf{e}},\phi _{\mathsf{e}} \right)$, we have $\rho_{_{\mathsf{P}}}\ne1$, which yields
			\begin{align}\label{C_rho<1}
				\mathsf{C}&=\log _2\left( 1+\frac{\alpha +\sqrt{\left( \overline{\gamma }G_{\mathsf{b}}\left( 1+\overline{\gamma }G_{\mathsf{e}}(1-\rho ) \right) +\overline{\gamma }G_{\mathsf{e}} \right) ^2}}{2(1+\overline{\gamma }G_{\mathsf{e}})} \right) \nonumber\\
				&=\log _2\left( 1+\frac{\overline{\gamma }G_{\mathsf{b}}\left( 1+\overline{\gamma }G_{\mathsf{e}}(1-\rho ) \right)}{1+\overline{\gamma}G_{\mathsf{e}}} \right).     
			\end{align}
			With ${\overline{\gamma}}\rightarrow\infty$, \eqref{C_rho<1} can be further approximated as follows:
			\begin{align}
				\mathsf{C}\approx \log _2\!\left( \frac{\overline{\gamma }_{\mathsf{b}}G_{\mathsf{b}}\overline{\gamma}_{\mathsf{e}}G_{\mathsf{e}}\left( 1-\rho \right)}{\overline{\gamma}_{\mathsf{e}}G_{\mathsf{e}}} \right)\! =\log _2\!\left( \overline{\gamma }_{\mathsf{b}}G_{\mathsf{b}}\left( 1\!-\!\rho \right) \right), 
			\end{align}
			which completes the proof of Corollary~\ref{cor_upw_p}.
			\subsection{Proof of Theorem \ref{USW_the}}\label{proof_usw}
			Based on \eqref{usw_model}, we can easily obtain the channel gains $G_k^{\mathsf{S}}$. We then focus on the channel correlation factor.
			By defining
			\begin{align}
				\delta_i=\left| \sum\nolimits_{m_i=-\tilde{M}_i}^{\tilde{M}_i}{\mathrm{e}^{\mathrm{j}\left( a_im_{i}^{2}+b_im_i \right)}} \right|^2
			\end{align}
			for $i\in \left\{ x,z \right\} $, the channel correlation factor for the USW model can be written as follows:
			\begin{align}
				\rho =\frac{\left| \mathbf{h}_{\mathsf{b}}^{\mathsf{H}}\mathbf{h}_{\mathsf{e}} \right|^2}{\lVert\mathbf{h}_{\mathsf{b}}\rVert^2\lVert\mathbf{h}_{\mathsf{e}}\rVert^2}
				=\frac{\left| \mathbf{h}_{\mathsf{b}}^{\mathsf{H}}\mathbf{h}_{\mathsf{e}} \right|^2}{G_{\mathsf{b}}^{\mathsf{S}}G_{\mathsf{e}}^{\mathsf{S}}}=&\frac{\delta_x\delta_z}{M^{2}}.
			\end{align}
			Since Bob and Eve are at different locations, $a_i$ and $b_i$ cannot be zero at the same time. To calculate $\delta_i$, we consider different cases as follows.
			\subsubsection{$a_i=0, b_i\ne0$}
			In this case, $\delta_i$ is given by
			\begin{align}
				\delta_i=\left| \sum\nolimits_{m_i=-\tilde{M}_i}^{\tilde{M}_i}{\mathrm{e}^{\mathrm{j}b_im_i}} \right|^2=\left| \frac{\sin \left( M_ib_i/2 \right)}{\sin \left( b_i/2 \right)} \right|^2.
			\end{align}
			\subsubsection{$a_i\ne0, b_i=0$}
			In this case, without loss of generality, we assume that $a_i<0$. By denoting $\vartheta  _i=a_iM_{i}^{2}$, $\delta_i$ can be calculated as follows:
			\begin{align}
				\!\!&\delta _i\!=M_{i}^{2}\left| \sum\nolimits_{m_i=-\tilde{M}_i}^{\tilde{M}_i}{\mathrm{e}^{\mathrm{j}\vartheta  _i\left( \frac{m_i}{M_i} \right) ^2}}\frac{1}{M_i} \right|^2\nonumber\\
				\!\!\!\!&\!\overset{\left( a \right)}{\approx}\!M_{i}^{2}\left| \int_{-\frac{1}{2}}^{\frac{1}{2}}\!{\mathrm{e}^{\mathrm{j}\vartheta  _ix^2}\mathrm{d}}x \right|^2\!\!\overset{\left( b \right)}{=}\frac{\pi}{\left| a_i \right|}\left| \mathrm{erf}\bigg( \frac{\mathrm{e}^{\mathrm{j}\frac{\pi}{4}}M_i\sqrt{\left| a_i \right|}}{2} \bigg) \right|^2\!\!,
			\end{align}
			where approximation $(a)$ follows from the concept of definite integral for large $T$, equality $(b)$ follows from the integral identity \cite[Eq. (2.33.3)]{integral}.
			\subsubsection{$a_i\ne0, b_i\ne0$}
			In this case, by denoting $\kappa _i=b_iM_i$ and $\nu  _i=\frac{a_i}{b_i}M_i$, $\delta_i$ can be calculated as follows:
			\begin{align}
				\delta _i&=M_{i}^{2}\left| \sum\nolimits_{m_i=-\tilde{M}_i}^{\tilde{M}_i}{\mathrm{e}^{\mathrm{j}\kappa _i\frac{m_i}{M_i}\left( 1+\nu _i\frac{m_i}{M_i} \right)}}\frac{1}{M_i} \right|^2\nonumber\\
				&\approx M_{i}^{2}\left| \int_{-\frac{1}{2}}^{\frac{1}{2}}{\mathrm{e}^{\mathrm{j}\kappa _ix\left( 1+\nu _ix \right)}\mathrm{d}}x \right|^2\nonumber\\
				&\overset{\left( a \right)}{=}\!\frac{\pi}{4\left| a_i \right|}\left| \mathrm{erf}\bigg( \frac{a_iM_i-b_i}{2\sqrt{\left| a_i \right|}}\mathrm{e}^{\mathrm{j}\frac{\pi}{4}} \bigg) +\mathrm{erf}\bigg( \frac{a_iM_i+b_i}{2\sqrt{\left| a_i \right|}}\mathrm{e}^{\mathrm{j}\frac{\pi}{4}} \bigg) \right|^2,
			\end{align}
			where $(a)$ follows from the integral identity \cite[Eq. (2.33.3)]{integral}.
			
			Combining the above cases, we obtain the results of \eqref{delta}, which completes the proof of Theorem~\ref{USW_the}. 
			\subsection{Proof of Theorem \ref{the_nusw}}\label{proof_nusw}
			Based on \eqref{distance} and \eqref{channel_NF}, the channel gain for user $k\in\{\mathrm{b},\mathrm{e}\}$ under the NUSW model can be written as follows:
			\begin{equation} \label{d1}
				\begin{split}
					G_k&=\frac{A\Psi _k}{4\pi r_k^{2}}\sum\nolimits_{m_x\in \mathcal{M}_x}\sum\nolimits_{m_z\in\mathcal{M}_z}\big((m_x^2+m_z^2)\varepsilon  _k^2\\
					&-2m_x\varepsilon  _k\Phi_k-2m_z\varepsilon  _k\Omega_k+1\big)^{-\frac{3}{2}}. 
				\end{split}
			\end{equation}
			We next define the function 
			\begin{align}
				f_k\left( x,z \right) \triangleq (x^2+z^2-2\Phi_kx-2\Omega_kz+1)^{-\frac{3}{2}}    
			\end{align}
			in the area $\mathcal{A} =\left\{ \left( x,z \right) \mid -\frac{M_i\varepsilon  _k}{2}\leq i\leq \frac{M_i\varepsilon  _k}{2},i\in \left\{ x,z \right\}  \right\}$ which is then partitioned into $M_xM_z$ sub-rectangles, each with equal area $\varepsilon  _k^2$. Since $\varepsilon  _k\ll 1$, we have $f_k\left( x,z \right) \approx f_k\left( m_x\varepsilon  _k,m_z\varepsilon  _k \right) $ for $\forall \left( x,z \right) \in \left\{ \left( x,z \right) \mid \left( m_i-\frac{1}{2} \right) \varepsilon_k\leq i\leq \left( m_i+\frac{1}{2} \right) \varepsilon  _k,i\in \left\{ x,z \right\} \right\}$. 
			
			Based on the concept of integral, we have 
			\begin{align}
				\sum\nolimits_{m_x,m_z}{f_k\left( m_x\varepsilon  _k,m_z\varepsilon  _k \right) \varepsilon  _k^2}\approx \iint_{\mathcal{A}}{f_k\left( x,z \right) \mathrm{d}x\mathrm{d}z}.   
			\end{align}
			As a result, \eqref{d1} can be rewritten as follows: 
			\begin{equation}\label{derivation}
				G_k=\frac{A \Psi _k}{4\pi d^2}\int_{-\frac{M_z\varepsilon  _k}{2}}^{\frac{M_z\varepsilon  _k}{2}}{\int_{-\frac{M_x\varepsilon  _k}{2}}^{\frac{M_x\varepsilon  _k}{2}}}f_k(x,z)\mathrm{d}x\mathrm{d}z.  
			\end{equation}
			We can calculate the inner integral with the aid of \cite[Eq. (2.264.5)]{integral} and the outer integral with the aid of \cite[Eq. (2.284.5)]{integral}, which yields the results of \eqref{Expression_Channel_Gain}. 
			
			Following similar steps to obtain \eqref{derivation}, the channel correlation factor can be written as follows:
			\begin{align}
				\rho &=\frac{A^2\Psi _{\mathsf{b}}\Psi _{\mathsf{e}}}{16\pi ^2G_{\mathsf{b}}^{\mathsf{N}}G_{\mathsf{e}}^{\mathsf{N}}r_{\mathsf{b}}^{2}r_{\mathsf{e}}^{2}}\nonumber\\
				&~~\times\left| \sum\nolimits_{m_x,m_z}{g_1\left( m_x\varepsilon _{\mathsf{b}},m_z\varepsilon _{\mathsf{b}} \right) g_2\left( m_x\varepsilon _{\mathsf{b}},m_z\varepsilon _{\mathsf{b}} \right)} \right|^2\nonumber\\
				&=\frac{A^2\Psi _{\mathsf{b}}\Psi _{\mathsf{e}}}{16\pi ^2G_{\mathsf{b}}^{\mathsf{N}}G_{\mathsf{e}}^{\mathsf{N}}r_{\mathsf{b}}^{2}r_{\mathsf{e}}^{2}\varepsilon _{\mathsf{b}}^{4}}\nonumber\\
				&~~\times\left| \int_{-\frac{M_z\varepsilon _{\mathsf{b}}}{2}}^{\frac{M_z\varepsilon _{\mathsf{b}}}{2}}{\int_{-\frac{M_x\varepsilon _{\mathsf{b}}}{2}}^{\frac{M_x\varepsilon _{\mathsf{b}}}{2}}{g_1\left( x,z \right) g_2\left( x,z \right) \mathrm{d}x}\mathrm{d}z} \right|^2.    
			\end{align}
			The above integrals can be numerically evaluated by utilizing the Chebyshev-Gauss quadrature rule, i.e., $\int_{-1}^1{\frac{f\left( x \right)}{\sqrt{1-x^2}}\mathrm{d}}x\approx \sum_{t=1}^T{f\left( x_j \right)}$ with $x_t=\cos \left( \frac{ 2t-1 }{2T} \pi \right) $, which leads to the expressions in \eqref{Expression_Squared_Correlation_Coefficient}. This completes the proof of Theorem~\ref{the_nusw}.
			
			\subsection{Proof of Corollary \ref{cor_ula}}\label{proof_ula}
				For the ULA case, the channel gains under the NUSW model can be written as follows:
				\begin{equation}\label{ula_gain}
					G_k=\frac{A\Psi _k}{4\pi r_k^{2}}\sum\nolimits_{m_z\in\mathcal{M}_z}(m_z^2\varepsilon  _k^2-2m_z\varepsilon  _k\Omega_k+1)^{-\frac{3}{2}}.
				\end{equation}
				We define the function $f'_k\left( z \right) \triangleq {( z^2-2\Omega_kz+1 ) ^{-\frac{3}{2}}}$
				over the line segment $\mathcal{L} =\left\{ z\mid -\frac{M\varepsilon _k}{2}\leqslant z\leqslant \frac{M\varepsilon _k}{2} \right\} $ that is then partitioned into $M$ subsegments, each with equal length $\varepsilon _k$. Since $\varepsilon _k\ll 1$, we have $f'_k\left( z \right) \approx f'_k\left( m_z\varepsilon _k \right)$ for $\forall z\in \left\{ z\mid \left( m_z-\frac{1}{2} \right) \varepsilon _k\leqslant z\leqslant \left( m_z+\frac{1}{2} \right) \varepsilon k \right\} $. Based on the concept of integral, we have $\sum_{m_z\in\mathcal{M}_z}{f'_k\left( m_z\varepsilon _k \right) \varepsilon_ k\approx \int_{\mathcal{L}}{f'\left( z \right) dz}}$. Consequently, \eqref{ula_gain} can be rewritten as follows:
				\begin{equation}\label{ula_gain_int}
					G_k=\frac{A \Psi _k}{4\pi d^2}\int_{-\frac{M\varepsilon  _k}{2}}^{\frac{M\varepsilon  _k}{2}}f'_k(z)\mathrm{d}z, 
				\end{equation}
				The integral can be calculated with the aid of \cite[Eq. (2.264.5)]{integral}, leading to the results of \eqref{ula_channelgain}. 
				
				Following the steps to obtain \eqref{ula_gain_int}, the channel correlation factor can be expressed as follows:
				\begin{equation}
					\rho =\frac{A^2\Psi _{\mathsf{b}}\Psi _{\mathsf{e}}}{16\pi ^2G_{\mathsf{b}}^{\mathsf{N}}G_{\mathsf{e}}^{\mathsf{N}}r_{\mathsf{b}}^{2}r_{\mathsf{e}}^{2}\varepsilon _{\mathsf{b}}^{4}}\left| \int_{-\frac{M\varepsilon _{\mathsf{b}}}{2}}^{\frac{M\varepsilon _{\mathsf{b}}}{2}}{{g'_1\left( z \right) g'_2\left( z \right) }\mathrm{d}z} \right|^2.    
				\end{equation}
				By applying the Chebyshev-Gauss quadrature rule to the above integral, we can obtain the results of 
				\eqref{ula_correlation}.

			\subsection{Proof of Lemma \ref{lem_depth}}\label{proof_lem_depth}
\begin{figure}[!t]
				\centering
				\setlength{\abovecaptionskip}{3pt}
				\includegraphics[height=0.25\textwidth]{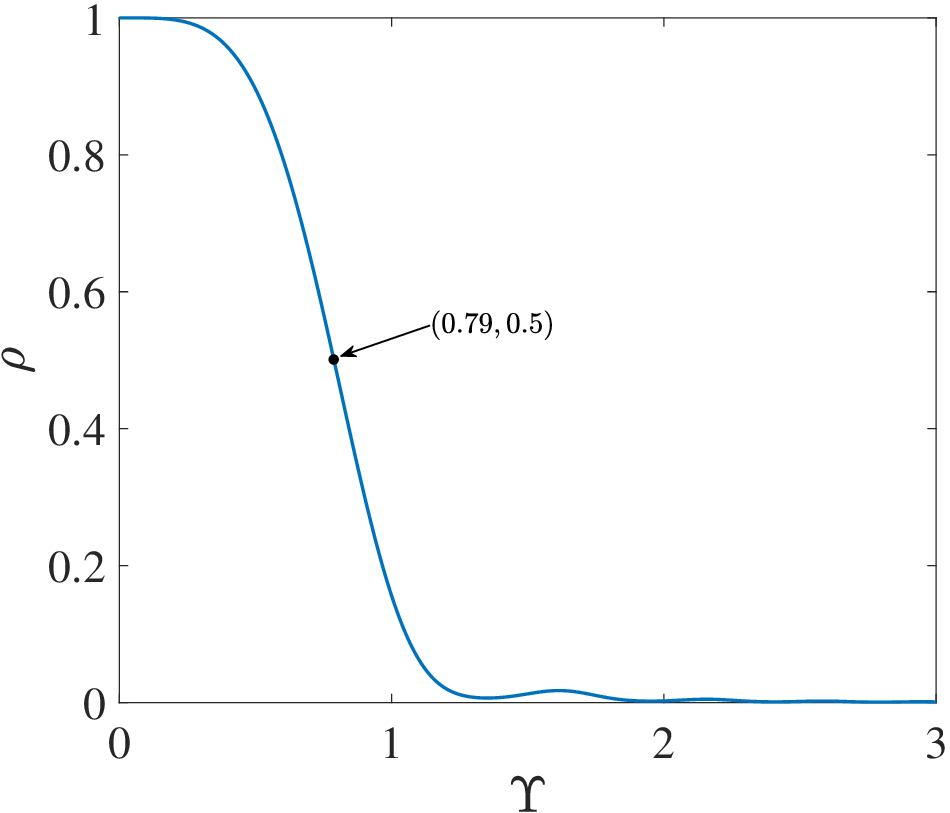}
				\caption{Numerical results of $\left| {\mathrm{erf}\left(  \sqrt{\pi}\mathrm{e}^{\mathrm{j}\frac{\pi}{4}}\Upsilon \right)}/{(2\Upsilon)} \right|^4$.}
				\label{num}
				\vspace{-5pt}
			\end{figure}

			Based on Lemma \ref{lem_angle}, we need to find the range of $r_\mathsf{e}$ such that $1-\rho\leq\frac{1}{2}$. From the results of Theorem \ref{USW_the}, when $M_x=M_z=\sqrt{M}$ and $(\theta_k,\phi_k)=(\frac{\pi}{2},\frac{\pi}{2})$, the channel correlation factor is given by
			\begin{align}
				\rho \triangleq\left| \frac{\mathrm{erf}\left(  \sqrt{\pi}\mathrm{e}^{\mathrm{j}\frac{\pi}{4}}\Upsilon \right)}{2\Upsilon} \right|^4,    
			\end{align}
			where $\Upsilon =\sqrt{\frac{Md^2}{4\lambda}| {r_{\mathsf{b}}^{-1}}-{r_{\mathsf{e}}^{-1}} |}$.
			It can be clearly shown from the numerical results given in Fig.~\ref{num} that $\left| \frac{\mathrm{erf}\left(  \sqrt{\pi}\mathrm{e}^{\mathrm{j}\frac{\pi}{4}}\Upsilon \right)}{2\Upsilon} \right|^4\geq \frac{1}{2}$ if and only if $\Upsilon \le 0.79$. Consequently, by defining $\Upsilon _{3\mathrm{dB}}\triangleq 0.79$, we have $1-\rho\leq\frac{1}{2}$ when 
			\begin{align}\label{upsilon_3db}
				\sqrt{\frac{Md^2}{4\lambda}\left| {r_{\mathsf{b}}^{-1}}-{r_{\mathsf{e}}^{-1}} \right|}\le \Upsilon _{3\mathrm{dB}}.    
			\end{align}
			Consequently, \eqref{upsilon_3db} is equivalent to
			\begin{align}
				\max \left\{ 0,{r_{\mathsf{b}}^{-1}}-{r_{\mathsf{s}}^{-1}} \right\} \le {r_{\mathsf{e}}^{-1}}\le {r_{\mathsf{b}}^{-1}}+{r_{\mathsf{s}}^{-1}}.
			\end{align}
			If $r_{\mathsf{b}}<r_{\mathsf{s}}$, the range of $r_{\mathsf{e}}$ satisfies
			\begin{align}
				r_{\mathsf{e}}\in \left[ \frac{r_{\mathsf{b}}r_{\mathrm{s}}}{r_{\mathsf{s}}+r_{\mathsf{b}}},\frac{r_{\mathsf{b}}r_{\mathrm{s}}}{r_{\mathsf{s}}-r_{\mathsf{b}}} \right], 
			\end{align}
			and thus the depth of insecurity is given by
			\begin{align}
				\mathcal{D} _{3\mathrm{dB}}=\frac{r_{\mathsf{b}}r_{\mathrm{s}}}{r_{\mathsf{s}}-r_{\mathsf{b}}}-\frac{r_{\mathsf{b}}r_{\mathrm{s}}}{r_{\mathsf{s}}+r_{\mathsf{b}}}=\frac{2r_{\mathsf{b}}^{2}r_{\mathrm{s}}}{r_{\mathsf{s}}^{2}-r_{\mathsf{b}}^{2}}.
			\end{align}
			If $r_{\mathsf{b}}\ge r_{\mathsf{s}}$, we have
			\begin{align}
				r_{\mathsf{e}}\in \left[ \frac{r_{\mathsf{b}}r_{\mathrm{s}}}{r_{\mathsf{s}}+r_{\mathsf{b}}},\infty \right],    
			\end{align}
			which yields $\mathcal{D} _{3\mathrm{dB}}=\infty $. This completes the proof.
			
			\subsection{Proof of Theorem \ref{theorem2}}\label{Appendix3}
			The eigenvalues of ${\bm\Theta}$ can be obtained from the characteristic equation as follows:
			\begin{align}\label{F1}
				\det(\mu{\mathbf I}-{\bm\Theta})=\det(\mu{\mathbf I}-\sigma _{\mathsf{b}}^{-2}{\mathbf h}_{\mathsf{b}}{\mathbf h}_{\mathsf{b}}^{\mathsf H}+2^{{\mathsf{R}_0}}\sigma _{\mathsf{e}}^{-2}{\mathbf h}_{\mathsf{e}}{\mathbf h}_{\mathsf{e}}^{\mathsf H})=0.
			\end{align}
			By defining $\bm \Phi\triangleq\mu{\mathbf I}-\sigma _{\mathsf{b}}^{-2}{\mathbf h}_{\mathsf{b}}{\mathbf h}_{\mathsf{b}}^{\mathsf H}$ and utilizing the matrix determinant lemma, \eqref{F1} can be rewritten as follows:
			\begin{align}\label{Ch_Poly2_1}
				(1+2^{{\mathsf{R}_0}}\sigma _{\mathsf{e}}^{-2}{\mathbf h}_{\mathsf{e}}^{\mathsf H}\bm\Phi^{-1}{\mathbf h}_{\mathsf{e}})\det(\bm\Phi)=0.
			\end{align}
			By employing the Woodbury matrix identity, we obtain
			\begin{align}\label{Ch_Poly2_2}
				\mathbf{\Phi }^{-1}= \frac{1}{\mu}\mathbf{I}+\frac{\sigma _{\mathsf{b}}^{-2}\mathbf{h}_{\mathsf{b}}\mathbf{h}_{\mathsf{b}}^{\mathsf{H}}}{\mu ^2-\mu \sigma _{\mathsf{b}}^{-2}\left\| \mathbf{h}_{\mathsf{b}} \right\| ^2}  .
			\end{align}
			Based on the Sylvester's determinant identity, we have
			\begin{align}\label{Ch_Poly2_3}
				\det(\mathbf{\Phi })=\mu ^{M-1}(\mu -\sigma _{\mathsf{b}}^{-2}\left\| \mathbf{h}_{\mathsf{b}} \right\| ^2).
			\end{align}
			Substituting \eqref{Ch_Poly2_2} and \eqref{Ch_Poly2_3} into \eqref{Ch_Poly2_1} yields 
			\begin{equation}\label{Ch_Poly2_4}
				\begin{split}
					\mu ^{M-2}&( \mu ^2+\left( 2^{\mathsf{R}_0}\sigma _{\mathsf{e}}^{-2}G_{\mathsf{e}}-\sigma _{\mathsf{b}}^{-2}G_{\mathsf{b}} \right) \mu \\
					&-\!2^{\mathsf{R}_0}\sigma _{\mathsf{b}}^{-2}\sigma _{\mathsf{e}}^{-2}G_{\mathsf{e}}G_{\mathsf{b}}\left( 1-\rho \right) ) =0.    
				\end{split}
			\end{equation}
			Let $\left\{\mu'_m\right\}_{m=1}^{M}$ denote the $M$ roots of \eqref{Ch_Poly2_4}. By the quadratic-root formula, we can obtain $\mu'_2=\ldots=\mu'_{M-1}=0$ and
			\begin{align}
				\mu' _1=\frac{\xi +\sqrt{\xi ^2+\chi}}{2},\ \mu' _M=\frac{\xi -\sqrt{\xi ^2+\chi}}{2}.
			\end{align}
			Since $\mu' _1\mu' _M=-\frac{\chi}{4}\le 0$, we have $\mu'_1\geq\mu'_2=\ldots=\mu'_{M-1}\geq\mu'_M$. Therefore, the principal eigenvalue of ${\bm\Theta}$ is $\mu'_1$, i.e., $\mu_{\bm \Theta}=\mu'_1$. The final results follow immediately.
			\subsection{Proof of Corollary \ref{cor_upw_minP}}\label{proof_cor_upw_minP}
			When Bob and Eve are located in the same direction, i.e., $\left( \theta _{\mathsf{b}},\phi _{\mathsf{b}} \right) =\left( \theta _{\mathsf{e}},\phi _{\mathsf{e}} \right)$, according to the results in Theorem \ref{UPW_the}, we have $\rho_{_{\mathsf{P}}}=1$ and $\frac{\sigma^{-2}G_{\mathsf{b}}^{\mathsf{P}}}{\sigma^{-2}G_{\mathsf{e}}^{\mathsf{P}}}=\frac{r_\mathsf{e}^2}{r_\mathsf{b}^2}$. In this case, if $r_\mathsf{e}\le2^{{\mathsf{R}_0}/2}r_\mathsf{b} $, we obtain $\mathsf{P}=\frac{2\left( 2^{\mathsf{R}_0}-1 \right) }{0}=\infty$. If $r_\mathsf{e}>2^{{\mathsf{R}_0}/2}r_\mathsf{b} $, $\mathsf{P}$ can be written as follows:
			\begin{align}
				\mathsf{P}=\frac{2^{\mathsf{R}_0}-1}{\sigma ^{-2}G_{\mathsf{b}}^{\mathsf{P}}-2^{\mathsf{R}_0}\sigma ^{-2}G_{\mathsf{e}}^{\mathsf{P}}}.
			\end{align}
			Since $\lim_{M\rightarrow\infty} G_k^\mathsf{P}=\infty$, we can obtain $\lim_{M\rightarrow\infty}\mathsf{P}=0$.
			
			When Bob and Eve are in different directions, we have $\lim_{M\rightarrow\infty} \rho_{_{\mathsf{P}}} = 0$, which yields
			\begin{align}\label{rho=0}
				\lim_{M\rightarrow\infty} \mathsf{P}&=\lim_{M\rightarrow\infty}\frac{2\left( 2^{\mathsf{R}_0}-1 \right)}{\xi+\sqrt{\left( \sigma ^{-2}G_{\mathsf{b}}^\mathsf{P}+2^{\mathsf{R}_0}\sigma ^{-2}G_{\mathsf{e}}^\mathsf{P} \right) ^2}}\notag\\
				&=\lim_{M\rightarrow\infty}\frac{2^{\mathsf{R}_0}-1}{\sigma ^{-2}G_{\mathsf{b}}^\mathsf{P}}=0.
			\end{align}
			The proof of Corollary~\ref{cor_upw_minP} is completed.
		\end{appendix}
		\bibliographystyle{IEEEtran}
		\bibliography{IEEEabrv}
		
\end{document}